\documentclass[graybox]{svmult}

\usepackage{mathptmx}       
\usepackage{helvet}         
\usepackage{courier}        
\usepackage{type1cm}        
\usepackage{makeidx}         
\usepackage{graphicx}        
\usepackage{multicol}        
\usepackage[bottom]{footmisc}
\usepackage{color}
\usepackage{enumerate}
\usepackage{amsmath}

\makeindex             

\usepackage{fancyhdr}
\begin{document}

\title*{Pure and Hybrid Evolutionary Computing in Global Optimization of Chemical Structures: from Atoms and Molecules to Clusters and Crystals}
\titlerunning{Pure and Hybrid Evolutionary Computing in Global Optimization}
\author{Kanchan Sarkar and S. P. Bhattacharyya }
\authorrunning{Pure and Hybrid Evolutionary Computing}
\institute{Kanchan Sarkar \at Department of Chemical Engineering and Materials Science, University of Minnesota, Minneapolis, Minnesota, USA \email{pcksarkar@gmail.com  \&  ksarkar@umn.edu}
\and S. P. Bhattacharyya \at Department of Chemistry, IIT Bombay, Powai: 400076, Mumbai, INDIA. \email{pcspb@iacs.res.in  \&   pcspb@chem.iitb.ac.in}}
%
%
\maketitle

\abstract{The growth of evolutionary computing (EC) methods in the exploration of complex potential energy landscapes of atomic and molecular clusters, as well as crystals over the last decade or so is reviewed. The trend of growth indicates that pure as well as hybrid evolutionary computing techniques in conjunction of DFT has been emerging as a powerful tool, although work on molecular clusters has been rather limited so far. Some attempts to solve the atomic/molecular Schrodinger Equation (SE) directly by genetic algorithms (GA) are available in literature. At the Born-Oppenheimer level of approximation GA-density methods appear to be a viable tool which could be more extensively explored in the coming years, specially in the context of designing molecules and materials with targeted properties.}
\tableofcontents
\section[Introduction]{Introduction} An N-atom system can be arranged in many different ways to form chemically meaningful structures. These structures represent minima on the 3N dimensional potential energy surface (PES) on which the nucleii move. One of these minima is the global one and search for the global minimum is a computationally challenging job. Optimization strategies play a very important role in theoretically elucidating the minimum energy structures supported on a potential energy surface. The ideas of PES and minimum energy structures are rooted in the Born Oppenheimer (BO) theory of molecules. BO theory starts by separating the slow nuclear motion from the much faster electronic motion which allows the electronic charge distribution in a molecule to adjust itself practically instantaneously as the nucleii move. The separation results in the creation of the so called adiabatic potential energy surface $E_n(x_1,x_2,\cdots , x_{3N})$ where $E_n$s are the eigenvalues of a molecular electronic Hamiltonian $H(r,x_1,x_2,\cdots , x_{3N})$ in which the nuclear displacement coordinates $(x_1,x_2,\cdots ,x_{3N})$ appear as parameters. The PES can have a number of stationary points. At all such points
\begin{eqnarray}
\frac{\partial E_n}{\partial x_i}=0, \text{ for } i=1,2, \cdots ,3N 
\end{eqnarray}
If in addition, the eigenvalues of the matrix of second derivatives of $E_n$ with respect to $3N$ nuclear displacement coordinates ($x_i)$ are $>0$, the stationary point is identified with a minimum energy structure. Out of $3N$ eigenvalues 3 representing transnational and 3 representing rotational motion of the center of mass are discarded as motion along these coordinates do not define any new chemical structures. It is immediately seen that the problem of locating minimum energy structures on an N-atom PES is a problem of minimizing a 3N dimensional function $E_n(X_1, X_2, \cdots, X_{3N})$. The function itself may be provided on the fly by solutions of the appropriate molecular Schrodinger equation or may be generated by superposing calibrated pair or higher body potentials.

In addition to the minima on the PES, chemists are also interested in locating first order saddle points and constructing minimum energy path (the reaction path) that connects the saddle point to two neighboring minima. At the first order saddle 
\begin{eqnarray}
\frac{\partial E_n}{\partial x_i}=0, \text{ for } i=1,2, \cdots ,3N 
\end{eqnarray}
and all but one of the eigenvalues (excluding the six) of the 2nd derivative matrix are positive, the special one being negative.

The present review focuses attention on the ways of exploring PES of molecules, clusters, crystals, etc. by invoking techniques of evolutionary computing for locating minima. In addition, we also review attempts to solve the atomic or molecular electronic Schrodinger equation (MESE) by similar techniques. In the case of molecular Schrodinger equation, the objective may be either to construct the PES by solving the MESE at a series of configuration and locating minima on the surface or to look for the deepest minimum in one go.

\section{The Optimization Problem} Any problem can in principle, be reduced to an equivalent optimization problem.  The domain of the function to be optimized is called the search space ($\mathbf{S}$). Goodness of any solution in $\mathbf{S}$ is measured with a mathematical function, called the objective function or fitness function ($\mathbf{O}:\mathbf{S}\rightarrow \mathbf{R}$). A fitness landscape (mapping from a configuration space into the real numbers) may be considered as a triple ($\mathbf{S},\mathbf{O},\mathbf{D}$), where $\mathbf{D}$ is a metric defined on $\mathbf{S}$.
 
In general, optimization problems involve setting a vector $\mathbf{X}$ of free parameters of a system in order to optimize (maximize or minimize) some objective function $\mathbf{O}(\mathbf{X})$ subject to the satisfaction of inequality constraints $g_k(\mathbf{X})$, equality constraints $h_k(\mathbf{X})$, as well as upper and lower bounds on the range of allowable parameter values.  Any constrained nonlinear optimization problem that deals with the search for a minimum of a nonlinear function $\mathbf{O}(\mathbf{X})$ of $m$ variables can be formulated as follows: 
\begin{eqnarray}
\min \mathbf{O}(\mathbf{X}), \; \mathbf{X} = (x_{1},\;x_{2},\;x_{3},\; \cdots,\;x_{m})^T\in \mathbf{S}\label{eqch1}
\\\quad\mbox{subject\; to}
\begin{cases}
g_{k}(\mathbf{X})=0,\; k=1,2,...,n\\
h_{k}(\mathbf{X})\leq 0,\; k=1,2,...,l\\
(x_{i})^{L}\leq x_{i} \leq (x_{i})^{U}
\end{cases}
\end{eqnarray}
Though equation (\ref{eqch1}) is defined as minimization problem, it can be equivalently posed as a maximization problem by a simple modification of the objective function, for example, by $-\:\mathbf{O}(\mathbf{X})$, $\frac{1} {\text{constant} + \mathbf{O}(\mathbf{X})}$ or by several other means. There are no specific conditions attached here to the variable type and the function characteristics may have multimodal fitness landscapes and significant levels of parameter interactions. Other conditions which are commonly addressed in optimization research but not specifically addressed in the present review include dynamic objective functions and multiple conflicting objectives. Let $\mathbf{X}\in \mathbf{S}$ be a feasible solution to the $m-$dimensional constrained nonlinear problem. $\mathbf{X}^\ast$ be the optimum vector that solves equation (\ref{eqch1}). The point $\mathbf{X}^\ast\in \mathbf{S}$ is a local optimal solution to this problem, if there exists an $\epsilon > 0$ such that
\begin{eqnarray}
\begin{cases}
\mathbf{O}(\mathbf{X})\ge  \mathbf{O}(\mathbf{X}^\ast),\: \forall \mathbf{X}\in \mathbf{S}\: : \: \parallel \mathbf{X}-\mathbf{X}^\ast\parallel <\epsilon \quad \text{for local minimum}  \\
\mathbf{O}(\mathbf{X})\le  \mathbf{O}(\mathbf{X}^\ast),\: \forall \mathbf{X}\in \mathbf{S}\: : \: \parallel \mathbf{X}-\mathbf{X}^\ast\parallel <\epsilon \quad \text{for local maximum}
\end{cases}
\end{eqnarray}
$\mathbf{X}^\ast\in \mathbf{S}$ is a global optimal solution if
\begin{eqnarray}
\begin{cases}
\mathbf{O}(\mathbf{X})\geq \mathbf{O}(\mathbf{X}^\ast),\: \forall \mathbf{X}\in \mathbf{S} \quad \text{for global minimum} \\
\mathbf{O}(\mathbf{X})\leq \mathbf{O}(\mathbf{X}^\ast),\: \forall \mathbf{X}\in \mathbf{S} \quad \text{for global maximum}
\end{cases}
\end{eqnarray}
The uniqueness of the global optimality is defined by
\begin{eqnarray}
\begin{cases}
\mathbf{O}(\mathbf{X})> \mathbf{O}(\mathbf{X}^\ast),\: \forall \mathbf{X}\in \mathbf{S} \quad \text{for global minimum} \\
\mathbf{O}(\mathbf{X})< \mathbf{O}(\mathbf{X}^\ast),\: \forall \mathbf{X}\in \mathbf{S} \quad \text{for global maximum}
\end{cases}
\end{eqnarray}
Global optimization methods differ from local optimization methods in that they attempt to find not just any local optimum, but the smallest (largest) local optimum in the search space $\mathbf{S}$. Global optimization is a difficult problem since no general criterion exists for determining whether the global optimum has been reached.

\section{Types of Optimization Algorithms}  
The goal of any optimization method is to design mathematical and computational infrastructure in order to locate the extremum (minimum or maximum) of a function (or functional). Automated optimization algorithm designing (i.e. the algorithm can learn adaptively to tune all its parameters by its own) can be very costly and not always straightforward. Expertise is needed to tackle with the parameter sensitivity, efficiency, robustness and solvability of the algorithms. Even mathematically equivalent formulations often differ substantially in efficiency and solvability. So it requires careful thoughts along with mathematical details while designing these optimization methods. Based on the use of random numbers, optimizers can be divided into two major categories, deterministic and stochastic. Deterministic algorithms are in general unidirectional (there exists at most one way to proceed, otherwise, the algorithm gets terminated) and do not use random numbers in any step of execution. State Space Search, Branch and Bound, Gradient based methods (such as Newton's method, Gauss-Newton method, Steepest-Descent method, Levenberg-Marquardt algorithms), Conjugate -- Direction methods (such as Conjugate-Gradient method, Fletcher-Reeves method, Powell's method, Partan method), Quasi-Newton methods (such as Davidon-Fletcher-Powell method, Broyden-Fletcher-Goldfarb-Shanno method, Hoshino method) are some examples of deterministic optimization methods \cite{C1_anton}. The gradient-based methods are the most used classical techniques for solving optimization problems. However, these methods can only be applied to the objective functions that are continuous and differentiable. Some of them, such as the Newton method, even require the knowledge of the second derivatives of objective functions as well. These techniques are also limited to the optimization problems having a sole optimum, as they are gradient-based. However, most of the objective functions in real-world problems are not differentiable and they have a large number of local optima, which makes these methods inapplicable. The unavailability of analytical gradients can be tackled by using numerical gradients. But this is only feasible for relatively low dimensional problems. So gradient-free optimization procedures are often much sought after techniques in fairly large scale optimization.

Stochastic algorithms, on the other hand, incorporate the concept of probability, employs at least one instruction or at least one operation that makes use of random numbers and do not use the gradient or Hessian matrix. The function to be optimized need not even be continuous or differentiable. If the relation between a candidate solution and its ``fitness" are not so obvious or too complicated, or the dimensionality of the search space is very high, it becomes harder to solve a problem deterministically. Apart from these reasons there can be many other pitfalls and booby traps \cite{C1_Deb,C1_Weis,C1_Weise,C1_Wei,C1_coel,C1_deb2,C1_Jin} (such as conflicting objectives, heavily constrained fitness function, oversimplification, overfitting, non-separability, scalability, rugged and deceptive fitness landscape, neutrality, epistasis etc.) that make the optimization problems difficult to handle and can lead an optimizer to a suboptimal region in the search space. Simulated annealing, Monte-Carlo sampling, Stochastic tunneling, Parallel tempering, Stochastic Hill Climbing, PSO, GA, Evolution Strategies, Memetic Algorithms, Differential Evolution are some examples of stochastic algorithms \cite{C1_Hol,C1_mich,C1_Bro,C1_tang,C1_Kirk,C1_Deb,C1_Weis,C1_Weise,C1_Wei,C1_talbi,C1_Jin,C1_coel,C1_deb2,C1_tk}. Our focus in this review has been mainly on pure evolutionary computing techniques and  their hybrids in which power of several methods are combined to produce a superior algorithm for global search. Averaged over the set of all problems, any pair of optimizers perform equivalently, which essentially implies that designing of a general purpose optimizer is bootless. Independent of the fitness function, one cannot (without prior domain knowledge) successfully choose between two algorithms based on their success in a different set of problems.

\section{Evolutionary Computation}
Biological life appeared on the earth probably as a culmination of many chance events involving chemical and physical interactions of molecules. Ever since the appearance of unicellular life, progressively complex life-forms have gradually evolved by the process of genetic evolution. The process of genetic evolution has elements of adaptive learning built into it. Evolutionary Computing methodologies are mathematical models of natural evolution implemented on a computer and are the most versatile complex problem-solvers. These techniques operate on a set of individuals or chromosomes (population) simultaneously. Each individual represents a potential solution to the problem being solved. The cardinality of population depends on the complexity of the problem. A chromosome is a sequence of genes (essentially the system parameters). Goodness of an individual is defined with a mathematical function, called fitness function ($f:S\rightarrow \mathbf{R}$). Individuals having lower fitness value are slowly washed out by the dominant competitors. There is thus a natural process of screening in the course of genetic evolution which finds expression in the Darwinian dictat of the survival of the fittest \cite{C1_darwin}. Chromosomal crossover and mutation produce new features in the chromosome. The evolution is a continuous process by which a species continuously strives to attain a genetic structure of the chromosomes that maximizes their probability of survival in a given environment. The evolutionary algorithms continue their iterations to improve the fitness of the individuals until an fitness maximal solution is found. The way individuals in the population are distributed has a major influence on the search. In the panmictic model, there is no structure in the population, all individuals are potential partners. Each individual can interact with every other one in the population during the evolutionary process. However, it is possible to define some structure providing the algorithm with higher exploration capabilities of the search space with respect to panmictic populations. There are two main canonical kinds of structured populations, namely distributed or coarse-grained and cellular or fine-grained \cite{C1_alba,C1_toma}. The choice must be made carefully.

\section*{Basic Ingredients of Evolutionary Computation}
In a diploid organism like the human being a pair of parental chromosomes form the chromosome-pair of the child, each chromosome containing millions of genes. One gene from the father and the corresponding gene from the mother constitute a gene-pair for the child, 
\begin{figure}[bth]
    \centering   \includegraphics[width=.76\linewidth]{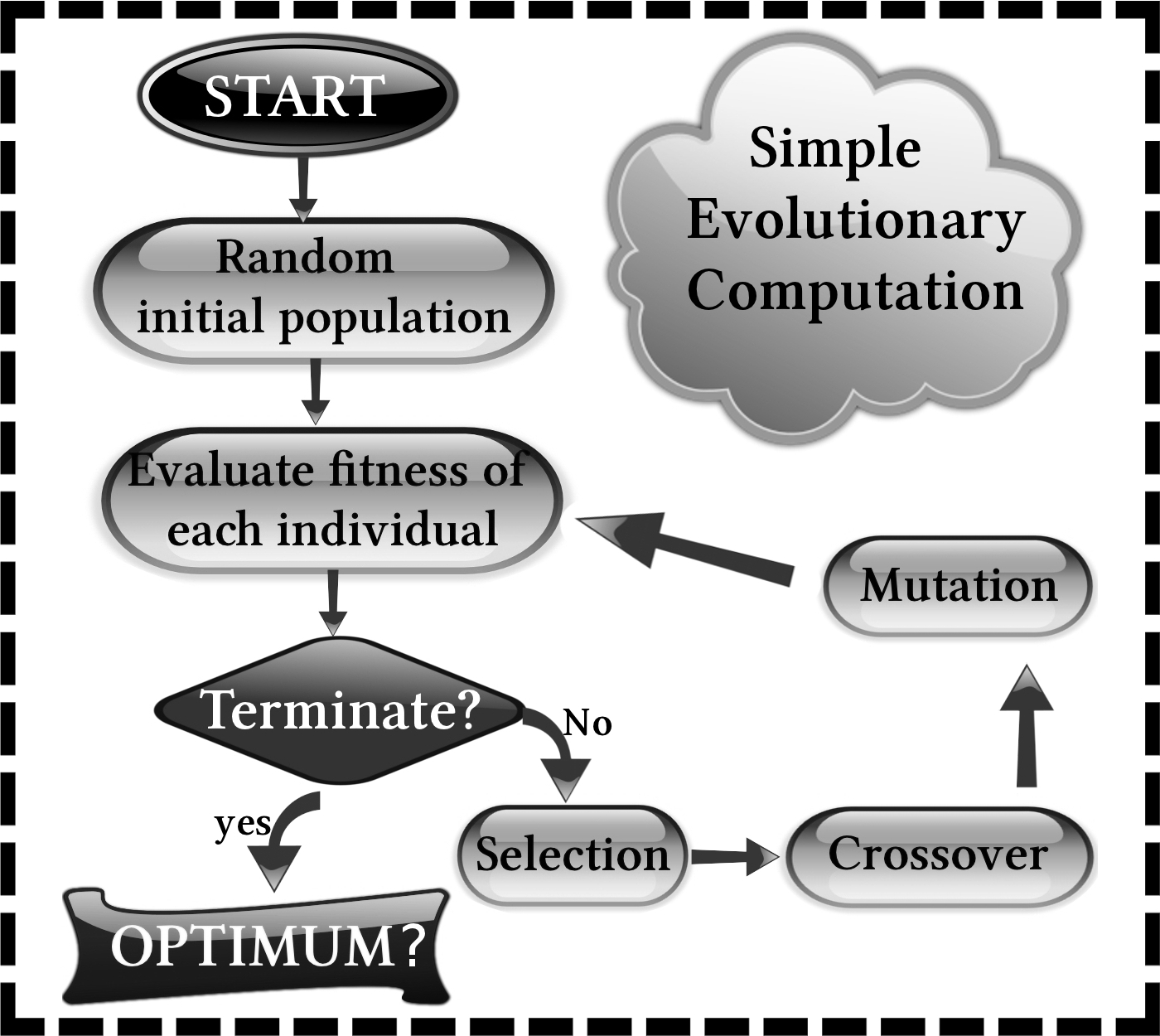}
      \caption[Flowchart for simple evolutionary computation]{Flowchart for simple evolutionary computation}\label{fig:EAFChrtnw}
\end{figure}
the gene-pair (genotype) determines the specific attributes or characteristics called the phenotypes. Most gene values (alleles) are inherited by the child from the parents unaltered; but on rare occasions, one or more of them may undergo change, by a process called mutation. If such changes are beneficial for the individual for better adaptation to the environment, they have higher probabilities of survival, and have correspondingly higher probabilities for producing their offsprings. Over generations those beneficial changes tend to stay on while the non-beneficial ones tend to disappear. In Evolutionary Computation, new individuals are generated in the search space by applying certain genetic operators to the current population. The dominant individuals will mate more often creating descendants with similar or better fitness in accordance with the natural selection process and statistically moving toward more optimal places in the search space. A simple picture of process steps in sequence for general Evolutionary Algorithms is shown in Figure \ref{fig:EAFChrtnw}. The three basic components of EC are A) Selection, B) Crossover, C) Mutation. Each component can be realized in a number of different ways. Let us start by looking at different ways of implementing a selection process on an evolving population of potential solutions

\subsection{Selection}Selection operator selects a proportion of the existing population with a specified probability to create the basis for the next generation. The operator is designed to ensure that promising solutions of the population have a greater probability of being selected for mating. The selection process is controlled by selection pressure, which is defined as the degree to which the better individuals are favored. So, selection process is essentially biased towards the more fit individuals and drives the Evolutionary Computation (EC) to improve the population fitness over the successive generations. Higher selection pressure can result in higher convergence rate, but with perhaps a higher chance of premature convergence. On the contrary, EC with low selection pressure will take unnecessarily longer time to find the optimal solution. In addition to the selection pressure, selection schemes should also try to preserve population diversity, to avoid premature convergence. The selection mechanism consists of two steps, the selection probability calculation and the sampling algorithm.
\subsubsection{Generational replacement:}Entire set of parents are replaced by their descendants or the $n$ worst parents are replaced with $n$ best offspring.
\subsubsection{Truncation Selection:} A proportion $(p)$ of the population are selected based on their fitness value and reproduced $\frac{1}{p}$ times so that the population size is maintained. Less fit individuals are not given the opportunity to evolve.
\subsubsection{Roulette Wheel Sampling Algorithm:}Individuals are selected according to their fitness values \cite{C1_Hol,C1_gold}. The higher the fitness, the higher the probability of being selected. For each individual $C_i$ in a population $P$ the selection probability ($p_s(C_i)$) is given by
\vspace{-0.2cm}\begin{eqnarray}
p_s(C_i)=\frac{f(C_i)}{\sum _{j=1}^{N}f(C_j)}\label{eq1}
\end{eqnarray}
The population is then mapped onto a roulette wheel, where each chromosome $C_i$ is given a slot that proportionally corresponds to $p_s(C_i)$. To select an individual, a random number is generated and the individual whose slot spans the random number is selected. The wheel is then spun $N$ times, $N$ being the cardinality of the population) to choose individuals until next generation mating pool is fully populated.
\subsubsection{Stochastic Universal Sampling:}Population is represented by a pie chart, in which each zone is allocated for an individual of the population. The areas of the zones are directly proportional to the fitness of the representative individuals, exactly as in roulette-wheel selection. But unlike roulette-wheel selection, here a single spin of the wheel is required to construct the mating pool. $N$ (cardinality of the population) number of equally spaced pointers are put around the pie chart. The position of first pointer is generated randomly in the interval ($0,1/N$). Individuals for mating pool are then selected by generating the $N$ pointers, starting with randomly generated first pointer, spaced by $1/N$, and selecting the individuals whose fitness spans the positions of the pointers. Stochastic Universal Sampling provides low spreading over the desired distribution of individuals and is bias-free. This approach is better than roulette wheel, because it keeps the diversity and prevents best individual from dominating the population.

\subsubsection{Rank-based Selection:}Individuals are ranked according to the fitness values in ascending order (least fit individual has rank=1, the fittest individual rank=$N$). Rank based selection uses a function to map the indices of individuals in the sorted list to their selection probabilities. The performance of the selection scheme depends greatly on this mapping function.

For linear rank-based selection, the bias is controlled by adjusting the selection pressure  ($SP$), such that $2.0\geq SP\geq1.0$, the expected sampling rate of the best individual being $SP$, while that of the worst individual is ($2-SP$). The selection pressure for all other individuals in the population can be obtained by linear interpolation of the selection pressure according to the rank. The fitness value for $i^{th}$ individual is calculated as: 
\begin{eqnarray}
F(i)=(2-SP)+2 \cdot (SP-1) \cdot \frac{r_i-1}{N-1}
\end{eqnarray}
where $F(i)\text{ and }r_i$ are the fitness and rank of the $i^{th}$ individual, respectively. Non-linear ranking permits higher selection pressures ($N-2\geq SP\geq1.0$) than what is used in the linear ranking method. The corresponding fitness values can be computed by using
\begin{eqnarray}
F(i)=\frac{N \cdot X^{r_i-1}}{\sum_{j=1}^{N}X^{j-1}} 
\end{eqnarray}
where $X$ is the root of the polynomial equation
\begin{eqnarray}
(SP-N) \cdot X^{N-1}+SP \cdot X^{N-2} +...+SP \cdot X +SP =0
\end{eqnarray}
Rank-based selection schemes help prevent premature convergence, but are computationally expensive because of the need to sort populations, and slow convergence.

\subsubsection{Boltzmann Selection:}A temperature like selection parameter controls this selection procedure. Initially the temperature has been kept high and then gradually lowered (selection pressure increases gradually) throughout the EA run. The probability of accepting an individual $k$:
\begin{eqnarray}
P(k\gets a)=\begin{cases}1,\quad \quad \quad \quad \: \quad if\: F_k\geq F_a \\
exp(\frac{F_k-F_a}{K_BT}),\quad if\: F_k<F_a \end{cases}
\end{eqnarray}
where $F_k\text{ and }F_a$ are the fitness of $k^{th}$ individual and average fitness of the population respectively. $K_B$ is Boltzmann constant and $T$ is the temperature.

\subsubsection{Tournament Selection:}A tournament is held among $n$ randomly picked competitors from the population. The individual with the best fitness value of those random $n$ tournament competitors (winner of the tournament) is then copied into the mating pool. A tournament size of $n=2$ is commonly used in practice. Selection pressure can be controlled by controlling the tournament size $n$. Tournament selection is superior to the proportionate selection as the latter is found to be significantly slower in terms of convergence time. Linear ranking and stochastic binary tournament selection have identical performance in expectation, but binary tournament selection is preferred because of its more efficient time complexity \cite{C1_Goldbe}. 
\subsubsection{Sexual Selection:} Crossover takes place only between chromosomes of opposite sex. The sex of the individuals in the current generation are determined either randomly or based on some specific criterion of the individuals. Generally all the females will get to reproduce only once regardless of their fitness level to facilitate exploration of the search space. The male selection is fitness biased. In feminine selection, the female chooses one of the males to mate based on a problem dependent attraction function \cite{C1_Goh,C1_Agraw}.

\subsubsection{Clonal Selection:}A proportion of the fitter individuals are selected for cloning. Each of these individuals receive a number of copies proportional to its position in the ranking. The clones then undergo the maturation process. A given individual and its maturated clones forms a subpopulation and the best of each subpopulation is allowed to pass to the next generation \cite{C1_Camp}.

\subsection{Crossover}
It is designed for sharing information between individuals by swapping or intermingling the genetic materials of two randomly chosen parent chromosomes, with the possibility that good chromosomes may generate promising descendants.
\begin{figure}[bth]
\includegraphics[width=.940\linewidth]{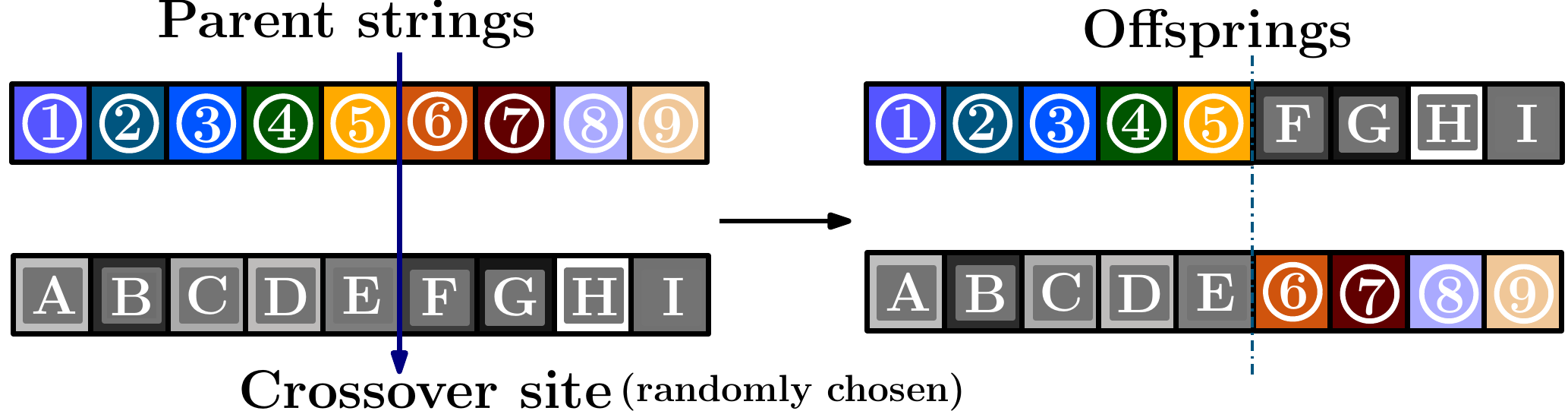}
\caption[Crossover]{Simple Crossover involves exchanging genetic information, beyond a randomly chosen cross site, by swapping the gene values within the parent's chromosome. Traditionally chromosomes are crossed at a single point. However, some problems could benefit from using multiple crossover points.}\label{fig:example5sc}
\end{figure}
A user defined crossover probability ($p_c$) determines the crossover frequency, i.e., how often crossover will take place. A general crossover scheme is shown in Figure \ref{fig:example5sc}. In the discussion of different crossover techniques that follows, we have represented the $j^{th}$ parent and offspring, of cardinality $d$, in a population as: \\
\begin{center}
\framebox{\begin{tabular}[t]{cccccc}
Parent &&&&& \quad Offspring  \\
$P_{j}(p_{1}^{j},p_{2}^{j},...,p_{d}^{j})$&&&&&\quad $O_{j}(o_{1}^{j},o_{2}^{j},...,o_{d}^{j})$ 
\end{tabular}}
\end{center}\vspace{0.2cm}

\subsubsection{Simple crossover:}A single crossover point $i\in \lbrace  2,\; 3,\; ...,\;d-1 \rbrace$ is  randomly chosen (say, i), where the two mating chromosomes are cut and the sections after the cuts are swapped to create two new individuals. $N$ point crossover is a generalization of this simple crossover applied to $N$ different segments.
\begin{eqnarray}
\begin{matrix} O_{ 1 }(o_{ 1 }^{ 1 },o_{ 2 }^{ 1 },...,o_{ d }^{ 1 })&=&(p_{ 1 }^{ j },\: p_{ 2 }^{ j },\: p_{ 3 }^{ j },\: ...,\: p_{ i }^{ j },_{ \searrow  }\: p_{ i+1 }^{ k },\: ...,\: p_{ d-1 }^{ k },\: p_{ d }^{ k }) \\ O_{ 2 }(o_{ 1 }^{ 2 },o_{ 2 }^{ 2 },...,o_{ d }^{ 2 })&=&(p_{ 1 }^{ k },\: p_{ 2 }^{ k },\: p_{ 3 }^{ k },\: ...,\: p_{ i }^{ k },^{ \nearrow  }\: p_{ i+1 }^{ j },\: ...,\: p_{ d-1 }^{ j },\: p_{ d }^{ j }) \end{matrix}\nonumber
\end{eqnarray}

\subsubsection{Cut and splice crossover:}This operator is an extension of simple crossover and designed specifically for cluster geometry optimization problems (see section Metal Clusters \& Nanoalloys).

\subsubsection{Uniform crossover:} It is assumed to be more disruptive in nature than simple crossover. So a lower crossover rate (say, 0.50) is used. Descendants are created by swapping the genes of two parent chromosomes with a predefined probability \cite{C1_Sys}.
\begin{eqnarray}
o_i&=&\begin{cases} p_{i}^{j},\quad if\quad r_i>0.5 \\ p_{i}^{k},\quad \text{otherwise} \end{cases}\\
r_i&\in &[0,1]\Longleftarrow \text{uniform random number}\nonumber\\
O_1(o_{1}^{1},o_{2}^{1},...,o_{d}^{1})&=&\left(p_{1}^{j},\: \boxed{p_{2}^{k}},\: \boxed{p_{3}^{k}},\: ...,\: p_{i}^{j},\:\, \boxed{p_{i+1}^{k}},\: ...,\: p_{d-1}^{j},\: \boxed{p_{d}^{k}}\: \right)\nonumber\\
O_2(o_{1}^{2},o_{2}^{2},...,o_{d}^{2})&=&\left(p_{1}^{k},\: \boxed{p_{2}^{j}},\: \boxed{p_{3}^{j}},\: ...,\: p_{i}^{k},\, \boxed{p_{i+1}^{j}},\: ...,\: p_{d-1}^{k},\: \boxed{p_{d}^{j}}\: \right)\nonumber
\end{eqnarray}

\subsubsection{Cell crossover:} Cells (a set of genes) of two random parent individuals are swapped with predefined probability to produce two descendants \cite{C1_iguo}.
\begin{scriptsize}
\begin{eqnarray}
\begin{matrix}  
{ P }_{ \searrow  \boxed { \diamondsuit ,\heartsuit ,\spadesuit ,\clubsuit , } \underbrace { \boxed { \heartsuit ,\spadesuit ,\diamondsuit ,\clubsuit , }  } \boxed { \diamondsuit ,\spadesuit ,\heartsuit ,\heartsuit , } \underbrace { \boxed { \heartsuit ,\clubsuit ,\clubsuit ,\diamondsuit , }  } \boxed { \diamondsuit ,\spadesuit ,\diamondsuit ,\spadesuit , } \boxed { \heartsuit ,\heartsuit ,\diamondsuit ,\clubsuit , } \boxed { \clubsuit ,\spadesuit ,\heartsuit ,\spadesuit , } \underbrace { \boxed { \spadesuit ,\clubsuit ,\clubsuit ,\diamondsuit  }  }  }^{ \nearrow \boxed { \spadesuit ,\clubsuit ,\diamondsuit ,\heartsuit , } \overbrace { \boxed { \clubsuit ,\diamondsuit ,\spadesuit ,\heartsuit , }  } \boxed { \spadesuit ,\diamondsuit ,\clubsuit ,\clubsuit , } \overbrace { \boxed { \clubsuit ,\heartsuit ,\heartsuit ,\spadesuit , }  } \boxed { \spadesuit ,\diamondsuit ,\spadesuit ,\diamondsuit , } \boxed { \clubsuit ,\clubsuit ,\spadesuit ,\heartsuit , } \boxed { \heartsuit ,\diamondsuit ,\clubsuit ,\diamondsuit , } \overbrace { \boxed { \diamondsuit ,\heartsuit ,\heartsuit ,\spadesuit  }  }  } \\ { O }_{ \searrow \boxed { \diamondsuit ,\heartsuit ,\spadesuit ,\clubsuit , } \boxed { \clubsuit ,\diamondsuit ,\spadesuit ,\heartsuit , } \boxed { \diamondsuit ,\spadesuit ,\heartsuit ,\heartsuit , } \boxed { \clubsuit ,\heartsuit ,\heartsuit ,\spadesuit , } \boxed { \diamondsuit ,\spadesuit ,\diamondsuit ,\spadesuit , } \boxed { \heartsuit ,\heartsuit ,\diamondsuit ,\clubsuit , } \boxed { \clubsuit ,\spadesuit ,\heartsuit ,\spadesuit , } \boxed { \diamondsuit ,\heartsuit ,\heartsuit ,\spadesuit  }  }^{ \nearrow \boxed { \spadesuit ,\clubsuit ,\diamondsuit ,\heartsuit , } \boxed { \heartsuit ,\spadesuit ,\diamondsuit ,\clubsuit , } \boxed { \spadesuit ,\diamondsuit ,\clubsuit ,\clubsuit , } \boxed { \heartsuit ,\clubsuit ,\clubsuit ,\diamondsuit , } \boxed { \spadesuit ,\diamondsuit ,\spadesuit ,\diamondsuit , } \boxed { \clubsuit ,\clubsuit ,\spadesuit ,\heartsuit , } \boxed { \heartsuit ,\diamondsuit ,\clubsuit ,\diamondsuit , } \boxed { \spadesuit ,\clubsuit ,\clubsuit ,\diamondsuit  }  } 
\end{matrix}\nonumber
\end{eqnarray}
\end{scriptsize}
\subsubsection{Heuristic crossover:}A single offspring is generated by linear extrapolation of two parent individual using equation \ref{coeq13} that may lie outside of the range of the two parent vectors \cite{C1_Wright}. It is also possible for this operator to generate an unfeasible offspring vector. In such a case another offspring created with different choice of $r$, $r$ being a uniform random variable belonging to the interval $[0, 1]$.
\begin{eqnarray}
\lbrace o_{i} \rbrace=\lbrace p_{i}^{j} + r (p_{i}^{j}-p_{i}^{k})\rbrace\;\;\;\;\; \mbox{ subject to } F(P_j)\geq F(P_k) \label{coeq13}
\end{eqnarray}
\subsubsection{Simplex crossover:}Simplex is constructed with randomly selected $N$ parents from the population. Centroid ($C$) of the simplex is calculated by
\begin{eqnarray}
C=\sum_{i=0}^{N}\frac{P_i}{N}\label{coeq16}
\end{eqnarray}
The simplex is then expanded by a small degree $\epsilon$
\begin{eqnarray}
O_j=C+\epsilon \cdot (P_j-C),\:j\in\{1,2,...,N\}\label{coeq17}
\end{eqnarray}
and within this expanded simplex, one or more new individuals are sampled \cite{C1_tsut}.
\begin{figure}[bth]
\includegraphics[width=.90\linewidth]{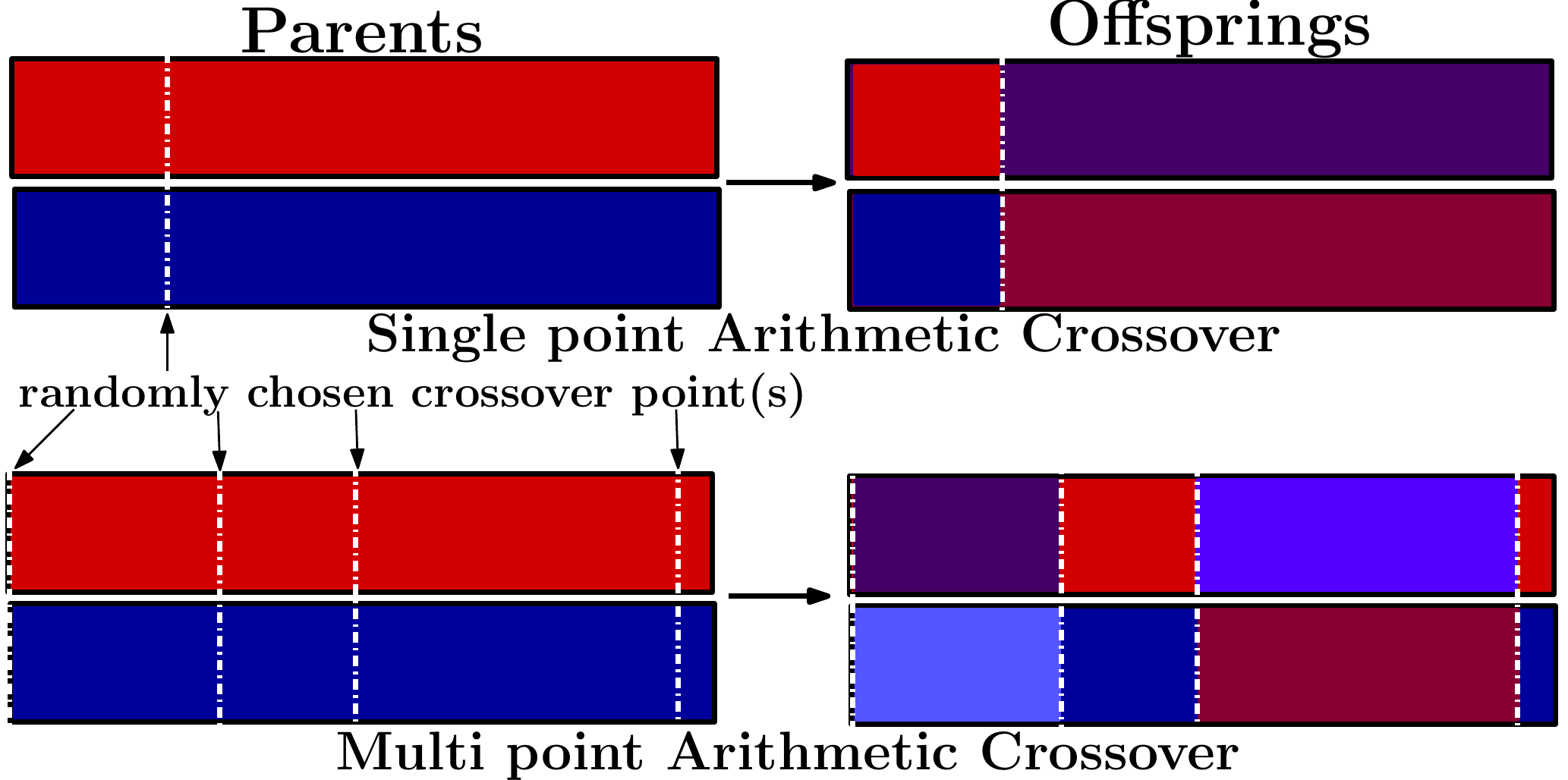}
\caption[Crossover]{Arithmatic Crossover involves intermingling genetic information, beyond a randomly chosen cross site and creates descendants that are the weighted arithmetic mean of two parents. For simplicity all the gene values in the parent strings are shown by same color.}\label{fig:example5}
\end{figure}
\subsubsection{Linear crossover:}Three offspring are generated (one in the exploitation zone and two in the exploration zone), with the two most promising solutions being chosen as descendants by means of fitness evaluation \cite{C1_Wright}.
\begin{eqnarray}
O_{1}= \frac{1}{2} P_{j} + \frac{1}{2}P_{k};\;\;\;\;\;\;
O_{2}= \frac{3}{2} P_{j} - \frac{1}{2}P_{k};\;\;\;\;\;\;
O_{3}= \frac{3}{2} P_{k} - \frac{1}{2}P_{j} 
\end{eqnarray}

\subsubsection{Arithmetic crossover:}Like simple crossover, chromosomes are crossed at a single or multiple crossover points. All positions arithmetic crossover are also useful. It was introduced to prevent invalid crossover and to increase searching space and convergence speed. Two complimentary linear combinations of the parents are generated on the basis of the arithmetic mean \cite{C1_michal}.
\begin{eqnarray}
\begin{matrix} O_1=\alpha \, P_k+(1-\alpha)P_j \\ O_2=\alpha\, P_j+(1-\alpha)P_k \end{matrix}\label{coeq18}
\end{eqnarray}
Arithmetic crossover operates in a way that every gene in descendants is a convex combination of the cooresponding genes from the two parents. $\alpha$ is constant in uniform arithmetical crossover or it may vary with generations in non-uniform arithmetical crossover. Figure \ref{fig:example5} schematically shows arithmatic crossover.

\subsubsection{BLX-$\alpha$  crossover:}To expand the range of arithmetic crossover Eshelman and Schaffer suggested blend crossover \cite{C1_Eshel}, which generates a single offspring by blending two floating point parent vectors as follows:
\begin{eqnarray}
\begin{matrix} o_{i}^{1}=\Re \left( (LB_i-\alpha \cdot I),(UB_i+\alpha \cdot I) \right)  \\ \text{where, } \begin{matrix} UB_i=\max (p_{i}^{j},p_{i}^{k}),\quad  & LB_i=\min (p_{i}^{j},p_{i}^{k}),\quad  & I=UB_i-LB_i \end{matrix} \end{matrix}
\end{eqnarray}
$\Re(a, b)$ is a function generating a uniform random number in the range $(a, b)$. The user-defined parameter $\alpha$ is usually set to a value of 0.5. Setting $\alpha=0.0 \text{ and } 0.25$ will give flat crossover \cite{C1_radc} and extended intermediate crossover \cite{C1_ein}, respectively.

\subsubsection{Similarity crossover:}In arithmetic crossover, there has always been a problem in selecting a perfect
weighting factor. The weight factor ($\alpha$) in arithmetic crossover (equation \ref{coeq18}) is replaced by similarity measurement $\left(S=e^{-EN}\right)$ between parents \cite{C1_Guve}
\begin{eqnarray}
EN=\frac{1}{\sqrt{d}}\cdot \left\lbrace (p_{1}^{l}-p_{1}^{m})^2+(p_{2}^{l}-p_{2}^{m})^2+ \cdots + (p_{d}^{l}-p_{d}^{m})^2 \right\rbrace^{\frac{1}{2}}
\end{eqnarray}

\noindent
In addition to these, there are other variants of crossover schemes such as parent-centric BLX-$\alpha$ crossover \cite{C1_manu}, Fuzzy recombination \cite{C1_voig}, SBX \cite{C1_Kaly}, PCX \cite{C1_Kalyan}, XLM \cite{C1_takah}, Laplace crossover \cite{C1_deep}, differential evolution \cite{C1_Storn}, partition \cite{C1_Whit,C1_Whitl}, linear BGA \cite{C1_Schl}, UNDX \cite{C1_ono}, fuzzy connectives based \cite{C1_herr}, direction-based \cite{C1_arum}, multiple crossover \cite{C1_chang}, Ring crossover \cite{C1_yil}, hybrid crossovers \cite{C1_gwiaz} are available in the literature.

\subsection{Mutation}The mutation operator perturbs one or more components (genes) of a selected chromosome, regulated by a predefined mutation probability, $p_m$, so as to increase entropy or to decrease the mutual information in the population. Mutation simply restores lost information or import unexplored genetic material into the population in order to distribute solutions widely across the search space and thus avoid premature convergence.

\begin{figure}[bth]
        \includegraphics[width=0.99\linewidth]{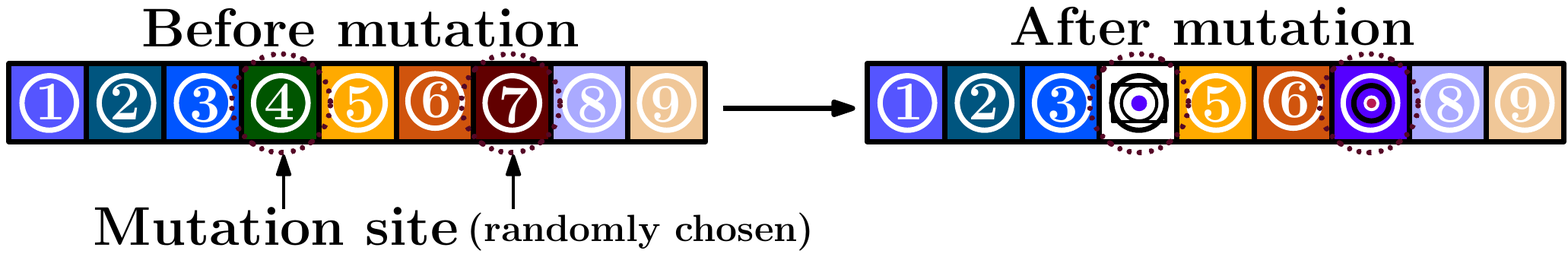}
        \caption[Mutation]{Mutation causes small alterations at randomly chosen gene(s) (with a predefined mutation probability) in an individual within available range of possible values.}\label{fig:example5dff}
\end{figure}

The performance of any evolutionary search algorithm heavily depends on the choice of its main control parameters: population size, mutation rate, and recombination rate and among these mutation rate is the most sensitive control parameter. Mutation-only-EA is possible, but, crossover-only-EA would not work with finite population size. There should be always a proper balance between exploration (discovering promising areas in the search space) and exploitation (optimizing within a promising area already found) abilities of a search algorithm. Crossover exploits existing genetic material, while mutation explores for newer material. Mutation does its best to avoid premature convergence and explore newer areas of search space. High mutation rate increases the probability of searching over large areas of search space; however, it prevents the population from converging to an optimum solution. On the other hand, too small a mutation rate may result in premature convergence. The best value of mutation rate is strongly problem specific and depends on the nature and implementation of the algorithm. There is no universal best mutation rate for most of the real world problems. The crossover and mutation rates (probability) are critical to the success of any EA, and are usually determined by means of trial-and-error. Automating the parameter settings i.e.,  making the search algorithm self-adaptive has been one of the most fashionable area of current interest among the EA community. A general mutation scheme is shown pictorially in Figure \ref{fig:example5dff}. We shall use specific notations for an individual before and after mutation, right through the discussions of different mutation schemes. The chromosome before and after mutation are expressed as follows:\\ 
\vspace{0.4cm}\framebox{\begin{tabular}[t]{cccccc}
Before mutation&&&&& After mutation \\
$X = ( x_1,x_2,\cdots , x_i,\cdots ,x_{d-1},x_d )$&&&&&$Y = ( y_1,y_2,\cdots , y_i,\cdots ,y_{d-1},y_d )$ 
\end{tabular}}

\subsubsection{Uniform mutation:}Genes in an individual are selected with a predefined probability ($p_m$) and replaced with a uniform random value according to \cite{C1_michal}.
\vspace{-0.3cm}\begin{eqnarray}
y_i=\begin{cases} \Re(LB_i,UB_i),{\quad \text{if}\: p_m>r_i},\quad r_i\in(0,1) \\ x_i,\: \quad  \text{otherwise} \end{cases}  
\end{eqnarray}
$\Re(a, b)$ is a function generating a uniform random number in the range $(a, b)$, $LB_i$ and $UB_i$ represents lower bound and upper bound to the $i^{th}$ gene. Uniform mutation tends to forestall premature convergence. In the neighborhood of the optimal region, however, uniform mutation rate may disrupt the fine-tuning of the search to the optimal point by inducing random oscillations.

\subsubsection{Nonuniform mutation:}Due to its dynamical nature, non-uniform mutation \cite{C1_michal} is one of the most commonly used mutation scheme in real coded EC. Gene(s), selected with a predefined mutation probability ($p_m$), are mutated as
\begin{eqnarray}
&\,&y_{ i }=\begin{cases} x_i+\Delta(t,UB_i-x_i),\: if\: r>0.5 \\ x_i-\Delta(t,\, x_i-LB_i),\: \text{otherwise} \end{cases}\\
&\,&\Delta(t,z)=z\left(1-r^{(1-\frac{t}{t_{max}})^b} \right)
\end{eqnarray}
where $r\in [0,1]$ is a random number, and $t$ is the generation number. The user defined parameter $b$ takes care of the non-uniform nature of the mutation step sizes.
\subsubsection{Gaussian mutation:}A noise with a mean of $0$ and a standard  deviation of 0.1 times  the  maximum value of the gene is added to the gene-value. A new individual is obtained by adding a random value to each element of the selected parent.
\begin{eqnarray}
y_i = \left \{
       \begin{array}{c}
           x_i+N(0,\sigma_i) \qquad  \text{Gaussian mutation \cite{C1_michal}}\\
           x_i+C(0,\tau_i) \qquad \; \text{Cauchy mutation \cite{C1_back}}\;\; \;   
         \end{array}
       \right.       
\end{eqnarray}
The mutation function that describes the transformation of an element $\vec{x}$ into
$\vec{y}$ \cite{C1_back}
\begin{eqnarray}
G = \left \{
       \begin{array}{c}
           \prod_{i=1}^{d}\frac{1}{\sqrt{2\pi}\sigma_i}e^{\left( -\frac{(\vec{y}_i-\vec{x}_i)^2}{2 \sigma_{i}^{2}} \right)} \qquad  \text{Gaussian}\\
           \pi^{-d}\prod_{i=1}^{d}\frac{\tau}{\tau^2+(\vec{y}_i-\vec{x}_i)^2} \qquad \;\;\;\;\;   \text{Cauchy}\;\; \;   
         \end{array}
       \right.       
\end{eqnarray}
$\tau$ is the scale parameter. L$ute{e}$vy-type mutation \cite{C1_iwam,C1_Tin}, Real Number Creep \cite{C1_davi} are some variants of Gaussian mutation. The scheme of Real Number Creep:
\begin{eqnarray}
y_i &=& \left \{
       \begin{array}{c}
           x_i+N(0,\sigma_{i}^{2}) \qquad \text{If } i=k\\
           x_i \qquad   \text{otherwise}\;\;\;\;\;\;\;\;\;\;\;\;\;
         \end{array} \qquad k=\text{Uniform}[1,d]
       \right.       
\\ \text{where, }\sigma{i}&=&\frac{UB_{i}-LB_{i}} {1000} 
\end{eqnarray}
Based on the PBX-$\alpha$ recombination operator and Gaussian distribution function, Dorronsoro et al. designed GPBX$-\alpha$ \cite{C1_dorr} mutation operator.
\subsubsection{Dynamic Random Mutation:} The scheme enables the mutation size to be adjusted dynamically \cite{C1_chen}.
\vspace{-0.2cm}\begin{eqnarray}
y_i = x_i +s \cdot \Phi_o\cdot (UB-LB)
\end{eqnarray}
where $\Phi_o$ is a random perturbation vector within the interval of $(0,1]$. The mutation step size is dynamically tuned by
\vspace{-0.4cm}\begin{eqnarray}
s=\left(1-\frac{t}{t_{max}}\right)^b
\end{eqnarray}
The parameter $b>0$ controls the decay rate of $s$ and $t_{max}$ denotes the maximum number of generations over which the search takes place. The parameter $b$ governs the shape of the allowable mutation region. The mutation range dynamically decreases as the number of generation increases. The idea is adopted from the annealing procedure in metallurgy. 
\subsubsection{Directed Random Mutation:}The value of a randomly selected gene is modified according to
\begin{eqnarray}
y_i = \left \{
       \begin{array}{c}
           x_i\pm \Delta_m \cdot r,\;\; \text{if}\;\; p_m<rand(0,1) \\
           x_i,\;\;\;\; \text{ otherwise} \;\;\;\;\;\;\;\;\;\;\;\;\;\;\;\;\;\;\;\;\;
         \end{array}
       \right.  
\end{eqnarray}
where $r$ is a random number in the interval $0<r<1$. The $+$ or $-$ sign is chosen with a probability of 0.5. $\Delta_m$ is the mutation intensity, dynamically adjusted on the basis of the degree of acceptability of mutation over a number of past generations \cite{C1_sharma,C1_sarkar}.
\begin{eqnarray}
\Delta_m = \left \{
       \begin{array}{c}
           \Delta_m \div (1+r_g),\;\; \text{if}\;\; N_{accept}<N_{lower}) \\
           \Delta_m \times (1+r_g),\;\; \text{if}\;\; N_{accept}>N_{upper})
         \end{array}
       \right.  
\end{eqnarray}
where $r_g$ is a Gaussian random number in the range ($0, 1$) and $N_{accept}$ is the number of accepted generation over past $N$ generations. $N_{lower}$ and $N_{upper}$ are the user defined lower and upper bound to $N_{accept}$.

\subsubsection{Wavelet mutation:}A continuous-time function $\psi(\omega)$ is called a ``mother wavelet" if it satisfies the following properties:
\begin{eqnarray}
&\;&\int_{-\infty}^{+\infty}\psi(\omega) d\omega=0\\
&\;&\int_{-\infty}^{+\infty}|\psi(\omega)|^2 d\omega<\infty
\end{eqnarray}
One example of such mother wavelet is Morlet wavelet:
\begin{eqnarray}
\psi(\omega)=e^{-\frac{\omega^2}{2}}cos(5\omega)
\end{eqnarray}
In order to control the magnitude and the position of $\psi(\omega)$ , $\psi_{a,b}(\omega)$ is defined as
\begin{eqnarray}
\psi_{a,b}(\omega)=\frac{1}{\sqrt{a}}\psi\left(\frac{\omega - b}{a}\right)
\end{eqnarray}
where $a$ is the dilation parameter and $b$ is the translation parameter. Ling et al. \cite{C1_ling}, based on the wavelet theory, proposed wavelet mutation  where each gene in a chromosome have a chance to mutate according to a preset mutation probabilty $p_m\in [0,1]$. The mutated gene is represented by
\begin{eqnarray}
y_i = \left \{
       \begin{array}{c}
           x_i+\psi_{a,o}(\omega) (UB-x_i)\qquad \text{if } \psi_{a,o}(\omega)>0\\
           x_i+\psi_{a,o}(\omega) (x_i-LB)\qquad \text{if } \psi_{a,o}(\omega)\leq 0
         \end{array}
       \right.       
\end{eqnarray}
For Morlet wavelet $\omega \in [-2.5,2.5]$ is randomly generated.
\subsubsection{BGA mutation:}The mutated gene $y_i$ is computed according to \cite{C1_ein}
\begin{eqnarray}
y_{i} = x_{i} \pm range_{i} \cdot \delta
\end{eqnarray}
The $+$ or $-$ sign is chosen with a probability of 0.5, $range_i$ is normally set to $0.1 \cdot (UB_i - LB_i)$ and $\delta$ is computed as follows.
\begin{eqnarray}
\delta=\sum_{i=1}^{d}\alpha_i 2^{-i} \qquad A_i \in 0,1
\end{eqnarray}
$\alpha_i$ is randomly generated and takes the value 1 with a probability $p_{\delta}=\frac{1}{d}$. The scheme ensures that at least one variable will be mutated in an individual. Discrete and continuous modal mutations \cite{C1_voigt} are generalization of BGA mutation. They only differ in how  $\delta$ is computed.

\subsection{Supplement to Selection: Elitism}
Sometimes crossover and mutation operations destroy potentially valuable genetic information which has evolved during the search. There is no guarantee whether EA will re-discover these lost information or not. If EA rediscovers those previously discarded information, it does so at the cost of some unnecessary time and computational power. To prevent this, the concept of elitism is introduced. Elitism ensure that a small proportion (typically 1 or 2) of the fittest individuals are passed onto the new generation so that the valuable information remains intact within the population and thus reduces the genetic drift. Higher proportion (causes rise in selection pressure) may lead to premature convergence. So the degree of elitism should be adjusted carefully. Elitism can speed up the performance of the GA significantly.

Apart from these operators, initial population also has major influence on the convergence of the algorithms. It is difficult to hit the global optimum in large systems starting from a fully randomized initial population as it leads to nearly identical glassy structures that have similar (high) energies and low degree of order \cite{C1_ognov}. Often random but symmetric structures may be helpful as initial guess for systems with large dimensions \cite{C1_Lyakho}. Presence of constraints often make an optimization problem difficult to solve. Several techniques such as adding penalty term to the objective function (the penalty increases with constraint violation), adjustment of the crossover or mutation operation, inducing instabilities to the infeasible solutions of the populations without modifying the objective function, have been proposed \cite{C1_Vl}.

\subsection{Evolutionary Computing Methodologies}\label{CH2PSO}
Let us consider several variants of EC methodologies that have been in use.
\subsubsection{Random Mutation Hill Climbing:} It is the simplest algorithm in the class of Evolutionary Computing methodologies. The algorithm works with one randomly generated string containing all the system parameters, (say, Cartesian position variables for geometry optimization) required for optimization and it involves mutation, as the only evolutionary process in order to climbing up the fitness landscape. Thus the underlying principle is to randomly choose a solution in the neighborhood of the current solution by means of mutating the string and retain this new solution if it improves the fitness function. A modified Random Mutation Hill Climbing was proposed and implemented by Sarkar et al. \cite{C1_sarkar,C1_ks1,C1_ks2,C1_ks3}. The modification lies in additional built-in features for enforcing adaptive control on all the parameters of the search heuristic. 

\subsubsection{Genetic Algorithms:} GAs are a class of stochastic heuristic search methods based on simplifications of evolutionary processes observed in Nature. They maintain a finite population size and are typically good at both the exploration and exploitation of the search space. They are invoked particularly in large, complex, and poorly understood problem domain. The search begins with randomly initializing a population of individuals followed by selection operation to create a mating pool. The mating process, implemented by crossing over the genetic material from two parents to form the genetic material for two new solutions. Random mutation(s) is(are) applied to promote diversity. If the new solutions are better than those in the population, the individuals in the population are replaced by the new solutions. The key idea is to maintain a population of chromosomes, that evolves over time through a process of competition and controlled variation. GA have had a great measure of success in search and optimization problems due to their ability to exploit the information accumulated about an initially unknown search space in order to bias subsequent searches into useful subspaces. The population scalability of performance in evolutionary computation has been an issue pursued since 1990s. The population sizing equation of Goldberg \cite{Gold}, M{\"u}hlenbein's study on minimal population size that can ensure convergence to the minimum \cite{bein}, the work relating to adaptive control of population size by Arabas et al \cite{abas} are the early examples of interest in this area. However, GA is trivially parallelized and its parallel efficiency could often outweigh the higher computational demand in every generation \cite{Nandys}.\label{GA_dis}

\subsubsection{Evolutionary Programming:}Both Evolutionary Programming (EP) and Evolution Strategy (ES) are known as phenotypic algorithms, whereas the GA are genotypic algorithms. Phenotypic algorithms operate directly on the parameters of the system itself, whereas genotypic algorithms operate on chromosomes or individuals representing the system. In recent work however the distinction has become blurred as ideas from the ES and EP have been applied also to strings or chromosomes that constitutes a population in GA. The primary difference between EP and the other approaches (GA, GP, and ES) is that no exchange of material between individuals in the population is made. Evolution is caused solely by appropriate mutation operators.

\subsubsection{Evolution Strategy:}The earliest form of ES is based on populations that consist of two competing individuals and the evolution process consists of the two operations mutation and selection. Only the fittest of the two individuals is allowed to produce descendants in the following generation. Later, general versions of the algorithm were developed through increasing the number of members in a population ($\mu$-parents and $\lambda$-descendants), among which the most common strategies are ($\mu, \lambda$) and ($\mu + \lambda$). In ($\mu, \lambda$) the parents are simply replaced with the descendants and in ($\mu + \lambda$), the parents compete with their children.

\subsubsection{Differential Evolution:}DE is a population based technique \cite{C1_ro,C1_in,C1_as}. The members of the population are chosen randomly to start with ($X_1,X_2,\cdots , X_n $). In a Differential Evolution (DE), mutation plays the dominant role. Mutation is used to produce new individual by adding the vector-difference between two randomly selected members (say, $X_k \text{ and }X_l$) of the population to another (say, $X_m$) producing a trial vector $V_i$, where,
\begin{eqnarray}
V_i=X_m+f(X_k - X_l)
\end{eqnarray}
$f$ being the mutation intensity. A total of $n$ such individuals are generated ($V_1$, $V_2$, $\cdots$, $V_n $). A crossover operator is then used to produce new candidate $U_i$ from the trial vector $V_i$ by setting the $j^{th}$ component of $V_i$ as follows:
\begin{eqnarray}
U_{ij}&=&V_{ij} \quad \text{if }r(j)\leq p_c \text{ or } j=j_{random} \\
      &=&X_{ij}, \text{otherwise}
\end{eqnarray}
The candidate is evaluated, and the better of the current individual and the candidate solution is selected.
\subsubsection{Genetic Programming:}GP is an extension of the GA in which the structures in the population are not fixed-length strings that encode candidate solutions to a problem, but programs expressed as syntax trees. So essentially Genetic Programming is a method to evolve computer programs. The genetic operators allow the syntax of resulting expressions to be preserved.
\subsubsection{Memetic Algorithms:}MAs are extensions of EA that repeatedly apply a local-search algorithm to exploit the search space in between generations to reduce the likelihood of the premature convergence.
\subsubsection{Particle Swarm Optimization:}PSO was introduced by Kennedy and Eberhart \cite{C1_Kenn} as a gradient free evolutionary computation technique that utilizes the `collective intelligence' of a cooperative swarm of particles. Each particle representing a candidate solution is randomly assigned variables to be optimized and then allowed to ``fly" with random velocities ($v_{min}\leq v \leq v_{max}$) in the search space. The task is to reach a point in the N-dimensional search space where the fitness has the maximum value and which is hopefully the optimal solution as well. Each particle keeps track of its coordinates in the search space, associated with the previous best fitness value ($p_{best}$). The overall best fitness value that the swarm has achieved so far is stored as $g_{best}$. At $t+1$ generation, velocity ($v$) and position ($x$) of $i^{th}$ particle are updated based on its current velocity ($v_{i}^{t}$), the distance from its previous best position (${p}_{i}^{t}$), and the distance from the global best position (${g}_{i}^{t}$) according to
\begin{eqnarray}
v_{i}^{t+1} &=& \omega{v}_{i}^{t}+C_1\phi_1\lbrace{p}_{i}^{t}-{x}_{i}^{t}\rbrace+C_2\phi_2 \lbrace{g}_{i}^{t}-{x}_{i}^{t}\rbrace \\
{x}_{i}^{t+1}&=&{x}_{i}^{t}+ {v}_{i}^{t+1}
\end{eqnarray}
The updating rule applies to all the components of velocities and positions. $C_1,C_2$ are the acceleration coefficients usually set to a value of 2.05, and $\phi_1,\phi_2$ are uniformly distributed random numbers within the interval of ($0,1$). The inertia weight $\omega$ decreases linearly from $\omega  = 0.9$ to $\omega  = 0.4$ over the whole run to control the momentum of the particles. The coefficient $C_1$ is called particle's self-confidence due to its a contribution towards the self-exploration of a particle, whereas $C_2$ is called swarm confidence due to its contribution towards motion of the particles in global direction, which takes into the collective wisdom \cite{C1_Das}. The velocities are assigned cutoffs $V_{min}$ and $V_{max}$ to prevent the swarm from disintegrating. The coordinates evolve continuously, and have no cutoffs.

\section{Monte Carlo Algorithms}
The main principles of Monte Carlo algorithm, named after a famous casino city in Monaco, are ergodicity and repeated random sampling to find the optimal solution. It was originally practiced under more generic names such as statistical sampling. In the 1940s, physicists working on nuclear weapons projects at the Los Alamos National Laboratory coined the present name. The random search, direct Monte Carlo sampling, the random hill climbing methods, Simulated Annealing, Quantum Annealing and Parallel Tempering are some examples that belongs to the class of Monte Carlo algorithms.

\paragraph*{Simulated Annealing:} The term ``annealing" is commonly used in metallurgy to define a process in which the metal is heated to a high temperature, followed by slow and controlled cooling. If molten metal is quickly cooled, it does not reach the perfect crystalline state of minimum free energy. The essence of the process lies in slow cooling which allows, time for redistribution of the atoms as they slowly lose mobility. Metropolis et al. \cite{C1_metro} proposed an algorithm for finding the equilibrium configuration of a collection of interacting atoms at a given temperature. The transition to a new state ($j$) from a state (i) is accepted with the probability 
\begin{eqnarray}
P(j\leftarrow i) = \left \{
       \begin{array}{c}
           1 \qquad \qquad \;\;\;\;\;\;\;\;\;\;\;\text{If } E_j\leq E_i\\
           exp\left(\frac{E_i-E_j}{K_BT} \right) \qquad \;  \text{otherwise}
         \end{array}
       \right.        \label{eq48}
\end{eqnarray}
Kirkpatrick et al. \cite{C1_Kirk} introduced the concept of Simulated annealing (SA) in combinatorial optimization following Metrpolis's idea. In SA, energy is regarded as the cost function and a set of parameters $\{x_{i}\}$ is used to define the configuration at some effective temperature. The target is to find the global minimum of the cost function using temperature as the control parameter. The SA first allows the system to `melt' at a high temperature and the temperature is lowered in small steps allowing it to spend sufficient time at each temperature until the system freezes. At each step of temperature, the sampling must be carried out for a sufficiently long time for the system to reach a steady state. The sequence of temperatures and the number of samplings allowed to reach `equilibrium' at each temperature constitute an ``annealing schedule". At each reconfiguring step perturbations are applied to generate a new point in the search space. The new point is accepted based on the Metropolis algorithm mentioned above (equation \ref{eq48}). Thus the algorithm differs from the Hill-Climbing method in the criterion used to replace the existing solution with a new one. The advantage of SA is the built-in local optimization scheme, since in the limit of the vanishing control parameter (T), the algorithm is a standard downhill simplex method. A performance comparison with EAs shows different strengths and weaknesses. While the short-range exploitation of SA based algorithms is generally considered to be better than that of pure EAs, EAs generally provide a better long-range exploration. Employing hybrid EAs with enabled local optimization steps in general makes the EAs at least on par with SA-based methods. 

An inverse approach (Reverse Monte Carlo) to the classical Monte Carlo algorithms successfully reproduces many salient
features of glass and metallic glass systems obtained from molecular dynamics simulation \cite{C1_Fang,C1_Chri}. Along with the success in different optimization problems SA has also been effectively used in quantum mechanical calculations, especially in the optimization of parameters of wave functions, constructing reaction paths for transformations on model potential energy surfaces and for rearrangement in Lennard Jones clusters \cite{C1_Biri,C1_Bi}.

\noindent
We have so far summarized the important aspects of evolutionary computing along with some relevant natural computing methods. We hope the readers will be able to exploit the information to form a well founded basis to choose appropriate schemes which maintain a good balance between exploration and exploitation for specific problems. In the next section we will provide an overview of applications of evolutionary computing and its hybrid methods in the general context of computing electronic structure of atoms, molecules and clusters.

\section{Evolutionary Computing in Electronic Structure Calculation}
The applications of EC to problems of calculation of electronic structure  of atoms, molecules, clusters and crystals have been interesting and numerous. We will review the applications under these categories:
\begin{itemize}
\item Solution of Schrodinger equation (SE) by EC.
\item Locating minima on the complex potential energy surfaces of molecules, clusters, crystal structure prediction.
\end{itemize}
\subsection{GA and Electronic Structure Calculation}
The first attempt to marshal the power of GA to tackle the problem of electronic structure calculation can be traced to a paper by Zeiri et al. \cite{C1_zeiri} on the application of GA to the calculation of bound state within the framework of a local density approximation  almost 30 years after GA came into being. GA was invoked for directly solving the time-independent SE \cite{C1_dhury1} by Chaudhury and Bhattacharyya in 1998. They tested the workability of their algorithm in finding the ground state of a particle moving in screened coulomb potential and that of an oscillator with quartic anharmonicity. They went on to examine the applicability of their method to the problem of solving the inhomogeneous differential equations of the RSPT with special reference to the ground states of two-electron atoms. Nakanishi and Sugawara (2000) proposed a ANN based way of solving the SE in which the wavefunction was represented on a perceptron type NN, and the weights and the biases were optimized by a micro-GA \cite{C1_Nak}.

Saha and Bhattacharyya (2001) proposed a stable and generalizable basis set free strategy of solving the SE by GA \cite{C1_Saha}. Test calculations on the ground and excited states of H-atom, coupled harmonic oscillators, and symmetric double well, demonstrated the robustness and workability of the algorithm proposed. Saha et al. (2003) parallelized \cite{C1_SaQ} the fitness evaluation steps and the parallel GA method was successfully invoked to solve an exactly solvable coupled oscillator problem and the H-atom problem. The kinetic energy was computed by numerical FFT while the potential energy was obtained by quadrature. Saha and Bhattacharyya (2004) further explored GA in solving the radial SE for the ground state of two electron atoms and ions \cite{C1_Sah}. The approach is basis set free and the results were better than HF results. The same authors extended the theory to calculate the ground state of Hydrogen molecule ion ($H_{2}^{+}$). In one realization, the internuclear distance (R) was allowed to evolve along with amplitudes of the electronic wave function at different points on a 3-d uniform grid. The equilibrium internuclear separation and the energy was obtained in a single run. Sahin et al. (2006) explored the GA based approach for computing the energy levels of hydrogenic impurity in a quantum dot and then went on to further extend the method for the self-consistent electronic structure calculation of many electron quantum dots \cite{C1_Sahi}.

GA has been deftly used to compute the molecular electronic structure of doped as well as undoped polythiophene and polyselenophene oligomers \cite{C1_sarkar,C1_pc1} within the framework of a modified SSH method. The density based approach in which the GA operators act only on the molecular geometry strings (arrays of the geometrical parameters) that uniquely define the corresponding fixed nuclei Hamiltonian $H(R)$. $H(R)$ in turn, defines a unitary transformation $U(R)$ which transforms a trial density $P(R_o)$ into a new density $P(R)=U(R)^{\dagger}P(R_o)U(R)$ at the new genetically modified geometry. Thus along with the geometry strings, the density $P(R)$ is forced to coevolve, till $[H(R),P(R)]=0$. The technique appears to be sufficiently flexible and could find use in ab-initio calculations of molecular electronic structure.

\subsection{GA in the Exploration of PES of Clusters}
Traditional optimization methods perform well on strongly convex response surfaces, whereas EAs are particularly appropriate for most of the real-world NP-hard problems in which the objective is multimodal and expensive to evaluate, the gradient of the objective function is not analytically definable, the problem domain is large in dimension and/or multiple linear and non-linear constraints have to be applied simultaneously. Traditional hard computing methods are occasionally fused with the evolutionary methodologies to develop computationally intelligent hybrid systems with moderate computational cost.

Atomic or molecular clusters are aggregates of similar or dissimilar  particles and generally have physical and chemical properties different from the bulk. Geometry optimization of atomic or molecular clusters belongs to the class of nondeterministic polynomial hard (NP-hard) problems. Exhaustive search throughout the PES in any reasonable amount of time is almost an impossible task. By the early 1990s EAs were applied to these classes of problems. Deaven et al. \cite{C1_Deav} applied GA to find fullerene structures up to $C_{60}$ starting from random atomic coordinates. Empirical Brenner potential was used to model the PES. Dugan et al. \cite{C1_Duga} had proposed Monte Carlo type local optimization between GA steps for moderately sized carbon clusters. Chaudhury et al. \cite{C1_Chau1} have used EA methods on empirical potentials to obtain optimized structures of halide ions in water clusters, which they then coupled with quantum-chemical (AM1) calculations for simulation of vibrational spectra. By suitably defining the objective functions, EC methods can also be used to locate additional features on the PES rather than just low-energy local and global minima. Chaudhury et al. \cite{C1_Chau2,C1_Chau3} have implemented methods for finding first-order saddle points and reaction paths, on PES of Lennard-Jones (LJ) clusters up to n=30.
               
Clusters may have properties which are different from those of discrete molecules or bulk matter: for example, some metals (e.g. palladium) which are non-magnetic in the solid state, are magnetic, with relatively high local magnetic moments in both neutral and anionic Pd$_N$ clusters \cite{C1_Mose}. The Coulomb explosion of large rare-gas clusters release huge energy (approaches the energy of nuclear processes). At very low temperatures ($< 2 K$ for $^4$He), He clusters may form a superfluid droplet. Metal nanoparticles play a crucial role as heterogeneous catalysts in petrochemical, pharmaceutical and clean energy sectors. Novel and interesting nanostructures, having useful chemical and physical properties could be obtained by controlling the nanoparticle size, composition, surface site preferences, and degree of segregation between the metal constituents \cite{C1_Nors}. Therefore, computing ground state configuration of atomic and molecular clusters is a challenging task from the theoretical perspective. The number of local minima of the clusters rises exponentially with the growth in cluster size. It increases further with composition heterogeneity because structures more complex than their homogeneous counterparts are possible. It is due to the existence of isomers with the same geometry and composition but showing different distributions of the constituent particles that the optimization task becomes notoriously difficult \cite{C1_Fend,C1_Bor,C1_Born,C1_Bornb,C1_cheng,C1_Huang,C1_Wilc,C1_Ferr,C1_ChenF}. For a cluster made of $n_A$ and $n_B$ atoms, with $n_A + n_B = N$, one finds that a single geometrical isomer can have $\frac{N!}{n_A! n_B!}$ different homotopes. Various attempts are there in the literature to find the global minimum energy structure of these systems, as for example the dynamic lattice searching method \cite{C1_Shao}, basin-hopping (BH) approach \cite{C1_Andr,C1_Ross,C1_kla,C1_Cerb}, adaptive immune optimization algorithm (AIOA) \cite{C_wux,C_wuxi,C_wuxia,c1_Genh} and evolutionary algorithms \cite{C1_Deav,C1_Pere1,C1_Pere2,C1_Pere3,C1_Pere4,C1_Marq,C1_Ferra,C1_Schw,C1_Mrq,C1_Cass,C1_Albe,C1_Fern,C1_Pull,C1_Oa, C1_h1,C1_h2,C1_h3,C1_h4}. Hybrid approaches combining evolutionary algorithm and a local search procedure, have been increasingly used to handle these problems. We review recent work in the area in what follows.

\subsubsection{LJ and Morse Clusters}
Ab initio quantum mechanical calculation for finding the global minimum geometry of a cluster of atoms or molecules is often extremely expensive \cite{C1_Alex,C1_Doll}. Computational cost increases with the accuracy of the calculation and size of the system. Generally empirical or semiempirical pair potentials are employed for large systems. LJ and Morse functions are the two most widely used and relatively simple pairwise additive potential models describing interactions between atoms which have been regularly used as benchmark to assess the performance of global search methods for cluster geometry optimization. The LJ and Morse potentials are respectively given by
\begin{eqnarray}
V_{ LJ }(r_{ij})&=&4\epsilon \sum _{ i=1}^{N-1} \sum _{j>i}^{N} \left[ \left( \frac{\sigma}{r_{ij}}\right)^{12}-\left( \frac{\sigma}{r_{ij}}\right)^{6} \right]\\
V_{M}(r_{ij})&=&\epsilon \sum _{i=1}^{N-1} \sum _{j>i}^{N} \left[e^{-2\beta (r_{ij}-r_{0})}- 2e^{-\beta (r_{ij}-r_{0})} \right]
\end{eqnarray} 
where $r_{ij}$ stands for the Euclidean distance between atoms i and j, $r_{0}$ is the equilibrium bond length, $\epsilon$ is the well depth of the pair-potential, $\sigma$ is the separation at which the pair-potential between the atoms goes through zero and $\beta$ is the interaction range scaling parameter. For the prediction of the ground state structure of crystals, the USPEX method 
\cite{C1_Glas,C1_Gao,C1_Oga1,C1_Oga2,C1_oganov,C1_Lyakhov,C1_Val,C1_ma,C1_Art,C1_w} has turned out to be extremely powerful and already guided materials scientists in finding interesting and unexpected crystal structures. Zhu et al. \cite{C1_Zhu} enhanced the efficiency of USPEX method by designing additional variation operators and constraints for partially or completely fixed molecules. Goedecker et al. \cite{C1_scho,C1_amsler,C1_si} offered improved minima hopping with a softening method and a stronger feedback mechanism to predict structures of homoatomic and binary clusters with LJ interaction as well as structures of silicon and gold clusters described by force fields. Cheng et al. \cite{C1_cheng} employed funnel hopping algorithm coupled with GA to locate the putative global minima of the LJ clusters and the Morse clusters up to N = 160. Wales et al. showed that the use of approximate symmetry provides a more productive way to explore the configuration space and substantially improves the efficacy of global optimization for the atomic clusters \cite{C1_Oakl,C1_Wales1,C1_Wales2,C1_Chakr}. Froltsov et al. \cite{C1_Fro} developed the ``cut and splice" GA augmented with twinning mutation moves for the structural optimization of both a series of single-funnel LJ clusters up to 70 atoms and the double-funnel LJ$_{38}$ cluster. Binary Lennard Jones (BLJ) cluster optimization is an even more challenging problem from the point of view of combinatorial complexity. They are also  interesting because catalytic properties of such clusters depend on the composition and structure. Comparatively little work has been done on the mixed clusters \cite{C1_souza,C1_Parod,C1_ue} as the search space is dramatically enlarged with the inclusion of more atom types. Mixed LJ clusters of widely varying compositions offer highly interesting additional perspectives on how variations in preferred structures emerge. Hybrid approaches like combining a global optimizer with a local search procedure have proved to be useful for mixed clusters \cite{C1_Marques,C1_Pereira,C1_Kolo,C1_Dor,C1_Diet,C1_Diete,C1_Dzhur}. 17 new putative global minima for BLJ clusters in the size range of 90-100 particles have been predicted by coupling hidden-force algorithm and non-Markovian parallel Monte Carlo search \cite{C1_Kolo}. Dor et al., proposed a multi-swarm based algorithm called PSO-2S \cite{C1_Dor} and tested it successfully on a number of benchmark functions. It uses charged particles in a partitioned search space and applies a repulsion heuristic on particles to increase the diversity for solving both unimodal and multimodal problems. Using GA-based global structure optimization framework and new set of fitted parameters for LJ potential from high-end ab initio calculations Hartke et al. optimized strongly mixed binary to quinary rare gas clusters \cite{C1_Diet,C1_Diete}. An average offspring method \cite{C1_scho} designed for cluster structure prediction was shown to perform better in systems with compact optimal structures. Two individuals are randomly selected and for each atom of the first cluster, the closest lying atom of the other cluster is identified. The corresponding atom of the child is placed randomly on the connecting line between the two parent atoms. The randomness of this operator is necessary to prevent the algorithm from producing a lot of identical offsprings. Leitao et al. \cite{C1_Leit} claimed that they applied for the first time an Island Model to the optimization of Morse clusters ranging from 41 to 80 atoms, combined with a hybrid steady-state evolutionary algorithm and a local optimization method. The performance was slightly more robust than that of the sequential approach. Dieterich et al. \cite{C1_Diete,C1_carsn,C1_han} designed EA program suite OGOLEM for structure optimization of mixed clusters. The OGOLEM framework provides both an MPI fronted for MPP parallelization and a threading fronted for SMP parallelization omitting unnecessary MPI overhead. They demonstrated the possibility to design molecules with targetted properties in the area of photochemistry using the OGOLEM framework \cite{C1_carsn}. There are prescriptions for the inclusion of Taboo-search features \cite{C1_Step,C1_Stepa} into the evolutionary algorithms which might help in reducing the amount of time spent in local optimizations rediscovering already known minima. Daskin et al. \cite{C1_skd} presented Group Leaders optimization algorithm in which the influence of the leaders in social groups is the inspiration for the evolutionary technique. The method is applied to locate the geometric structures of LJ clusters as well as to the quantum circuit design problems. Deep et al. \cite{C1_eep} made an attempt to solve LJ problem by incorporating a multi-orbit dynamic neighborhood topology in PSO.  In multi-orbit topology, the swarm has heterogeneous connectivity with some subsets of the swarm strongly connected while with the others are relatively isolated. This heterogeneity of connections balances the exploration-exploitation trade-off in the swarm. The  dynamic neighborhoods topology helps avoid entrapment in local optima. 

\subsubsection{Ionic clusters}
Oleksy et al. \cite{C1_Olek} calculated the equilibrium geometries and dissociation energies of electronic ground-state of $He_{N}^{+}$ clusters (N = 3-35) via an extended GA method and employing a semiempirical valence-bond model of intracluster interactions. They found that for the cluster sizes of N = 3, 4, 7-9, and 14-35, the positive charge of $He_{N}^{+}$ delocalizes over a trimer ionic core, which is more or less linear and centrosymmetric. For N = 2, 10-13, a dimer ionic core develops, while for N = 5 the positive charge is delocalized over the whole cluster; for N = 6, a tetramer ionic core is formed. Atoms outside the ionic core are neutral and are distributed among several solvation rings. The electronic and geometrical structures of sodium cluster anions \cite{C1_Hub} [Na$_{n}^{-}$, n=20-57] and cationic Na$_{n}^{+}$ clusters \cite{C1_Man} were determined by applying GA -- density functional theory (DFT). Structures of Indium oxide nanoclusters and zirconia nanoclusters have been predicted by a method combining a robust evolutionary algorithm with classical interatomic potential and quantum chemical models \cite{C1_walsh,C1_Wood}. Kim et al. \cite{C1_Kim} used GA to design Nanoporous TiO$_2$ for Low-Temperature Processable Photoanodes of Dye-Sensitized Solar Cells. Darwinian and Lamarckian schemes within evolutionary algorithms have been compared in the context of structure prediction of the titania (TiO$_2$) polymorphs. Lamarckism in natural evolution regards the effects of ``inheritance of acquired characters" as the motive force of evolution, while the Darwinism claims that evolution is nothing but the cumulative processes of natural selection with random mutation and denies the possibility of inheritance of acquired characters. Although the mainstream of today's evolutionary theory follows Darwinism, it has been found that the Lamarckian scheme is more successful and efficient at generating the target structures \cite{C1_Woodley}. A combination of Buckingham and LJ potential functions describe the interactions between different atomic species in the system
\begin{eqnarray}
V(r_{ij})=A_{ij}e^{-r_{ij}/\rho_{ij}}+\frac{B_{ij}}{r_{ij}^{12}} -\frac{C_{ij}}{r_{ij}^{6}}  
\end{eqnarray} 
The parameters $A_{ij}$, $B_{ij}$, $C_{ij}$, and $\rho_{ij}$ are dependent on the species involved in the interaction. To perform a global geometry optimization of clusters resulting from the microsolvation of alkali metal ions (i.e., Na$^{+}$, K$^{+}$, and Cs$^{+}$) with benzene molecules Marques et al. \cite{C1_Mard} used modified LJ function for nonelectrostatic contributions. Zhu et al. \cite{C1_Zu} explored all the possible stoichiometries for Mg-O system at pressures up to 850 GPa. They found that two extraordinary compounds MgO$_2$ and Mg$_3$O$_2$ became thermodynamically stable at 116 GPa and 500 GPa, respectively. They predicted the existence of thermodynamically stable Xe-O compounds \cite{C1_zhuty} at high pressures (XeO, XeO$_{2}$ and XeO$_{3}$ became stable at pressures above 83, 102 and 114 GPa, respectively). Woodley et al. \cite{C1_Al} predicted a greater stability of tetrahedral and trigonal coordinations compared to the tetragonal one for zinc oxide clusters [(ZnO)$_n$, n=1-32] using an evolutionary algorithm with polarizable shell interatomic potentials. Wang et al. \cite{C1_WangP} successfully applied PSO for the prediction of elemental (Li, C, Mg, Si), binary (SiO$_{2}$, SiC, ZnO, TiH$_{2}$, TiB$_{2}$, MoB$_{2}$), and ternary (MgSiO$_{3}$ and CaCO$_{3}$) compounds in various chemical-bonding environments (metallic, ionic, and covalent bonding). Li et al. \cite{C1_Lir} obtained the geometrical structures of (Al$_{2}$O$_{3}$)$_{n}$ (n = 1-7) clusters via GA coupled with the DFT. Avaltroni et al. \cite{C1_Av} identified peculiar low energy isomers of small clusters (C$_{2}$Al$_{4}$ and CB$_{6}^{2-}$) using both standard random search and GA procedures. Pal et al. \cite{C1_Pal,C1_pal2} determined the lowest energy structure of ZnS quantum dots of different sizes by using a search based on GA coupled with the density-functional tight-binding method (DFTB) and found a new ring-like configurations of ZnS quantum dots which had higher HOMO-LUMO gaps compared to other ZnS quantum dot structures. Pereira et al. developed an evolutionary algorithm for the global minimum energy search for water clusters up to (H$_{2}$O)$_{20}$, benzene clusters up to (C$_{6}$H$_{6}$)$_{30}$ and (C$_{6}$H$_{6}$)$_{n}^{+}$ with n=2-20 \cite{C1_Lla}. Do et al. \cite{C1_dom} identified the energy minima of water, methanol, water + methanol, protonated water, and protonated water + methanol clusters with DFT combined with basin hopping. Addicoat et al. \cite{C1_ddi} optimized a GA in order to identify the minimum energy structure of arbitrarily hydrogenated and hydroxylated fullerenes, and implemented the method for exhaustive calculations on all possible isomers. They suggested that crossover operator did not have  significant effect on the search efficiency of GA and hence the most efficient version of their GA does not employ crossover, thereby reducing it to an Evolutionary Algorithm (EA). Cai et al. \cite{C1_Cai} used GA to optimize structures of silicon clusters. Using the GA, Zhang et al. \cite{C1_Zhan} predicted the most stable structures and a number of low-energy metastable structures for Si[001] symmetric tilted grain boundaries with various tilt angles which are found to be in very good agreement with the results of first-principles calculations. Yao et al. \cite{C1_Ya} investigated the high-pressure structures of solid nitrogen through GA combined with first-principles electronic structure calculations. Hooper et al. \cite{C1_oop} studied generic defect association complexes in metal oxide materials at experimentally relevant dopant concentrations by a specialized GA-inspired search procedure for doped metal oxides. The search algorithms had been tested on lanthanide-doped ceria (L = Sm, Gd, Lu) with various dopant concentrations. 

\subsubsection{Metal Clusters \& Nanoalloys}
Chen et al. \cite{C1_Ch} proposed a parallel differential evolution for cluster optimization (PDECO) with triangle mutation and migration operators and applied it to Pt clusters with great efficiency. Rogan et al. \cite{C1_Jo} implemented a GA based methody on small Pd Clusters. The lowest-lying isomers of the copper cluster, Cu$_{9}$, have been obtained by combining a GA driven approach with DFT \cite{C1_Ass}. Assadollahzadeh et al. \cite{C1_s} employed a seeded GA technique using DFT together with a relativistic pseudo-potential to search for global and energetically low-lying minimum energy structures of neutral gold clusters Au$_{n}$ (n = 2-20). GA coupled with a tight-binding potential is employed by Ping et al.\cite{C1_dLi} to optimize neutral lead clusters Pb$_{n}$ (n = 2-20). Pereira et al. \cite{C1_PereFra} made a study on the effectiveness of different crossover operators in the global optimization of atomic clusters. They came up with a `Cut and Splice' crossover operator which proved to be extremely useful. The first parent string is ``Cut" randomly along a random horizontal cutting plane into two complementary parts and the second parent string is ``Cut" in such a way that the number of the atoms beneath and above the cutting plane are equal to the number of atoms of the two complementary parts of the first parent. ``Splice" joins the head of one
\begin{figure}[bth]
\includegraphics[width=.999\linewidth]{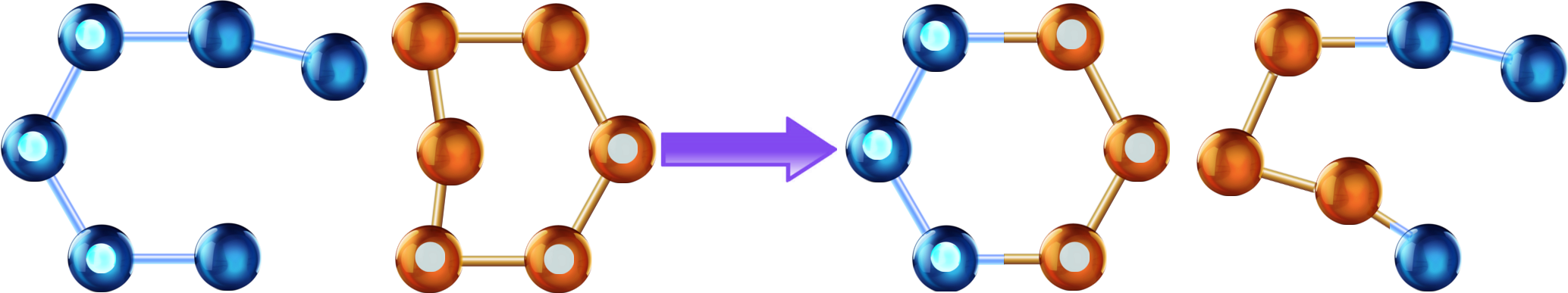}
        \put(-300,35){\rotatebox{0}{$\LARGE\mathbf{P}_1$}}
        \put(-225,35){\rotatebox{0}{$\LARGE\mathbf{P}_2$}} 
        \put(-125,35){\rotatebox{0}{$\LARGE\mathbf{O}_1$}}
        \put(-50,38){\rotatebox{0}{$\LARGE\mathbf{O}_2$}}                   
        \caption[Crossover]{To make the operation consistent and produce descendants with same number of atoms: the center of mass of the parent structures have to be translated to the origin of the coordinate system and have to rotate the plane of the parent clusters until the generated offspring contains the correct number of atoms.}\label{fig:xample5}
\end{figure}
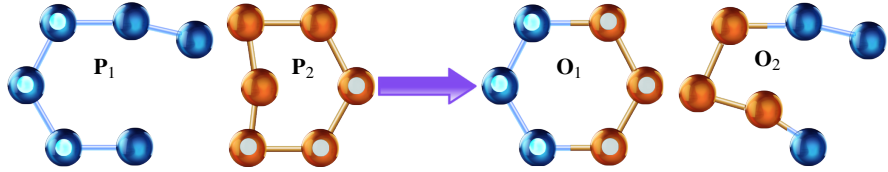
string with the tail of the other one. Pereira et al. \cite{C1_PereFra} modified this operator by the way it determines the sub-clusters to be exchanged. Here a random atom from the parent string is selected and placed on the offspring. A random number ($M\in [1,N-2]$) is generated (N is the number of atoms in the cluster). M atoms from the first parent string closer to the first selected atom are chosen and placed on the offspring. The remaining (N-M-1) atoms are taken from another parent string closer to the first selected atom. Those atoms which are already occupied are skipped. Figure \ref{fig:xample5} schematically depicts `Cut and splice crossover'. The energetic ground states of gold clusters, for some magic numbers, with up to 318 atoms were obtained by minima hopping method \cite{C1_Bao}. A GA--DFT was invoked by Sai et al. \cite{C1_Sai} who explored the size evolution of structural and electronic properties of neutral gallium clusters of 20-40 atoms. The atomic structures and electronic states of Ga$_{n}$ clusters significantly differ from the solid but resemble solid and liquid to a certain extent. Meng et al. \cite{C1_Meng} found that NbCl$_{5}$ and ZnMg intercalation in graphite results in p- and n-type doping, respectively. To model clusters they used Gupta semiempirical potential and obtained globally optimal structures of those metal clusters employing GA with ``cut-and-splice" crossover and triangle mutation. The optimized clusters were then placed between two layers of graphite, each of which contained 12 $\times$ 12 primitive cell and 288 atoms. The whole system was optimized with Universal potential and the electronic structures at the optimized doped cluster geometries were finally calculated using DFT. The atomic arrangement has an immense contribution to the catalytic activities of a nanoalloy. So, the development of a reliable method for predicting atomic arrangements has become increasingly important. Hong et al. \cite{C1_Hong} combined GA with first-principles electronic structure calculations to find the most stable configurations of elementary Au$_{m}$ and Ag$_{n}$ clusters, as well as binary clusters Au$_{m}$Ag$_{n}$ ($5 \leq m + n \leq 12$). The search led to the critical Au$:$Ag ratios for which the 2D-3D transition takes place. The study showed that Ag atoms prefer peripheral positions with lower coordination number while Au atoms tend to occupy central sites with higher coordination numbers. Johnston et al. \cite{C1_Hei} investigated small Sn-Bi clusters employing a DFT/GA method. Oh et al. \cite{C1_Oh} developed a combination of GA and classical molecular dynamics (MD) simulations to evaluate the optimal arrangement of multilayered core-shell structure for the Pt-Cu nano alloy with 1,654 atoms, regardless of the composition ratio. Using Tabu search in descriptor space and DFT Orel et al. \cite{C1_orel} found the global minima of the neutral binary Sn$_{m}$Pb$_{n}$ atomic clusters, $7 \leq m + n \leq 12$, for all the possible stoichiometric ratios. Fournier et al. \cite{C1_Fo} found the minimum-energy structures of Ag$_{n}$Rb$_{n}$ (n = 2-10) clusters by a combination of DFT and Tabu search. Chu et al. \cite{C1_Chun} described a new fragment-based evolutionary algorithm (EA) for de novo optimization, specifically developed to handle organometallic and transition metal compounds. Au atoms preferentially segregates on the surface of Pd-Au nanoalloys. DFT calculations on fcc-type cubo-octahedral Pd-Au nanoparticles have indicated that the surface Pd atoms occupy (111) rather than (100) faces, thereby maximizing the number of relatively strong surface Pd-Au bonds \cite{C1_Yuan}. Larger cohesive energy, lower electronegativity and smaller atomic radius (minimize bulk elastic strain) of Pd favors a Pd core and Au surface shell by lowering surface energy of the cluster. Besides electron transfer from Pd to more electro-negative Au, greater order of the Pd-Au than the Pd-Pd and Au-Au bonds favors Pd-Au mixing \cite{C1_Bornb}. Ferrando et al. had considered a methodology based on extensive global-optimization within empirical potential models and subsequent DFT based local relaxation of low-energy structures pertaining to different structural motifs (or basins on the energy landscape) of gas-phase alloy nanoclusters \cite{C1_Ferra,C1_Pazb,C1_Epit,C1_Barc,C1_Pazbr,C1_Paz,C1_and} modelled by the many-body Gupta potential.  Pittaway et al. \cite{C1_Bornb} performed global optimization of Pd-Au bimetallic clusters using a GA, coupled with the Gupta many-body empirical potential (EP) to model inter-atomic interactions. Johnston and coworkers \cite{C1_Bor,C1_Born,C1_Ferr,C1_ChenF,C1_Ferra,C1_Pazbr,C1_Paz,C1_Nu,C1_r1,C1_r2,C1_r3,C1_r4,C1_r5,C1_ei,C1_Br} did a series of development on theoretical simulation of alloys and bimetallic clusters (e.g., Cu-Ag, Ni-Ag, Cu-Au, Ag-Pd, Ag-Au, Au-Pd) with GA coupled with the Gupta many-body potential. $Ag_{core}Pd_{shell}$ nano catalysts could be important in the development of micro polymer electrolyte membrane fuel cells for portable devices, and could also be applied to the promotion of other catalytic reactions \cite{C1_Kara}. $Ag_{3}Pd_{10}$ cluster was identified as a wedge-shaped nano-shell with $C_s$ point group symmetry by a combination of GA global optimization and DFT calculation. Wu et al. \cite{C_wuxi} performed Global optimization of $Ag_mPd_n (m + n = 15)$ and $Ag_{3m}Pd_{38-3m}(m = 1-12)$ clusters and came to the conclusion that silver atoms had a strong tendency towards segregating at the surface. Combined experimental/theoretical studies have indicated the preference for $Pd_{core}Au_{shell} \;(\sim 5nm)$ and an ``onion-like" $Pd-Au-Pd$ configuration for larger $Pd-Au$ nanoalloys $ (\sim12 nm)$. Johnston et al. \cite{C1_Born} have studied mixed coinage metal clusters, using the Gupta potential.

\subsection{Crystal Structure Prediction}
Evolutionary algorithms \cite{C1_Lyakho,C1_Art,C1_w,C1_Lyakhov,C1_Gao,C1_Oga2,C1_oganov,C1_Zhu,C1_scho,C1_Olek,C1_walsh,
C1_Wood,C1_Woodley,C1_WangP,C1_Ogary,C1_oganorn,C1_Bahmann,C1_hu,C1_Ciob,C1_ham} have provided reliable and powerful means of exploring the PES of crystals leading to the identification of the most stable, interesting and sometimes unexpected crystal structures. There are several other methods such as basin hopping \cite{C1_Ji,C1_Waleoga,C1_G,C1_scho,C1_amsler}, metadynamics \cite{C1_ZhuC,C1_Qiang}, PSO \cite{C1_WangP,C1_Wang}, and simulated annealing \cite{C1_Do} for performing similar search. For large dimensional problems, due to the complex nature of PES and the presence of high-energy disordered structures (random initial population) these algorithms often encounter serious challenges. Lyakhov et al. introduced an additional variation operator - the coordinate mutation or soft mutation \cite{C1_Lyakho} for the purpose of crystal structure prediction. Instead of using complete randomness in the mutation operator concerted mutation which directs the system to choices that have a higher probability to improve the fitness of the solution string is introduced. The idea of soft-mode mutation or softmutation \cite{C1_Lyakho,C1_Lyakhov} is to move the atoms along the softest modes i.e., the lowest frequency eigenmodes that correspond to directions of lowest curvatures of the energy surface. The eigenvector corresponding to lowest non-zero eigenvalue determine the direction of softmutation. To calculate the softest modes one has to construct the dynamical matrix. It is often enough to have an approximate direction and sufficiently large mutation amplitude to arrive at a new low-energy structure. There is no need therefore of doing computationally expensive ab initio dynamical matrix calculation. Cheaper method for example dynamical matrix computed from bond hardness coefficients can be used. The soft mutation favor atoms with higher local order,  Lyakhov et al. \cite{C1_Lyakhov} constructed the first generation using pseudo-subcells with fractional atomic occupancies to improve the order and diversity of the structures. They used a fingerprint function that improves the selection process through removing clones. Superconducting high pressure phase of germane \cite{C1_Gao}, structural characterization of compressed silane \cite{C1_Martin}, rhombohedral structure of superhard BC$_2$N \cite{C1_oli}, superconducting structures of BC$_5$ \cite{C1_li}, superhard monoclinic polymorph of carbon \cite{C1_QuanP}, high-pressure orthorhombic polymorph of MgB$_{2}$ stable above 190 GPa \cite{C1_Yanm}, high-pressure phases of CaLi$_{2}$ \cite{C1_XiePh}, metallic structures of oxygen at pressures in the range of 100-250 GPa \cite{C1_Oga2}, two new hexagonal ultra-hard phases of $WN_2$ \cite{C1_WangB}, two unique high-pressure metallic phases of Stannane (SnH$_{4}$) at superconducting temperatures of 15-22 K for the Ama2 phase at 120 GPa and 52-62 K for the P6$_{3}/$mmc phase at 200 GPa \cite{C1_GaoZ} and SiH$_{4}$(H$_{2}$)$_{2}$ at high superconducting temperatures of 98-107 K at 250 GPa \cite{C1_LiYin,C1_GaoZ} are some of the systems explored through ab initio evolutionary methodology. By combining PSO with first principles calculations, Xiang et al. \cite{C1_ng} discovered a new metastable Si phase (Si$_{20}$-T has a quasidirect gap of 1.55 eV) which could be a promising solar energy absorber. Oganov et al. \cite{C1_oganov} found a new boron phase (comprised of icosahedral B$_{12}$ clusters and B$_{2}$ pairs in a NaCl-type arrangement), stable between 19 and 89 GPa, and exhibiting evidence for charge transfer. In the context of reliably predicting ground state geometry, the Universal Structure Predictor: Evolutionary Xtallography (USPEX) \cite{C1_Glas,C1_Art,C1_w,C1_Zhu,C1_o7} method has turned out to be extremely successful. Oganov et al. using USPEX methodology first demonstrated the possibility of computing hardness (at least for insulators and semiconductors) just from the crystal structure opens up the possibility of global optimization of hardness, aimed at the computational discovery of new superhard materials. They had introduced local measures of the quality of structure to locate defective regions in the crystal and used fingerprint niching, soft-mode mutation, symmetry- and pseudosymmetry-enabled generation of structures to greatly speed up the search for the global minimum \cite{C1_o1,C1_o2}. In a recent article \cite{C1_o3} they have summarized the different applications of the USPEX method as a tool for crystal structure prediction and showed that this method has an enormous applications in both computational materials design and studies of matter at extreme conditions. In a recent application Oganov et al. showed in conjunction with the first-principles calculations, evolutionary algorithm will lead to the discovery of novel dielectrics \cite{C1_o4}. In another application they successfully predicted energetically lower novel 2D boron structure than the $\alpha$-sheet structure \cite{C1_o5}, stable hafnium carbides \cite{C1_o6}. Woodley et al. predicted ground state geometry for a series of transparent conducting indium oxide \cite{C1_walsh} and zirconia nanoclusters \cite{C1_Wood} by combining evolutionary algorithm with classical interatomic potential and quantum chemical models. Wu et al. showed that structure exploration by classical potentials with the accuracy of density functional theory is an  efficient scheme for complex crystal structure prediction \cite{C1_qw}. In another work they favorably compared Lamarckian schemes within evolutionary algorithms with Darwinian schemes in the context of minimum energy structure prediction of titania phases \cite{C1_Woodley}. O{'}Keeffe \cite{C1_Keef} describes several difficulties encounted in the crystal structure prediction. Study on GA and minima hopping method based crystal structure prediction reveals that for relatively smaller sizes both methods show comparable efficiency while for larger systems GA becomes advantageous over minima hopping \cite{C1_Ji}. Oganov et al. explained how and why evolutionary crystal structure prediction works the way they do \cite{C1_nov}. Using adaptive GA Zhao et al. \cite{C1_xzao1} studied the structures and stabilities of the alkaline earth metal peroxide XO$_2$. They also predicted complex crystal structures of the orthorhombic, rhombohedral, and hexagonal polymorphs close to the Zr$_2$Co$_{11}$ intermetallic compound \cite{C1_xzao}.  Li et al. \cite{C1_L} explored the high-pressure crystal structures of Mg by the PSO algorithm. Wang et al. have developed a software package titled Crystal Structure Analysis by Particle Swarm Optimization (CALYPSO), enforcing symmetry constraints on structure generation, bond characterization matrix for elimination of similar structures, introducing partial random structures per generation for enhancing structural diversity, and penalty function, etc. \cite{C1_WangP,C1_Wang} for predicting crystal structure from random initial starting geometries. They have applied CALYPSO to crystal structure prediction, earth and planetary materials \cite{C1_m1,C1_m2,C1_m3,C1_m4,C1_m5,C1_m6,C1_m7,C1_m8}. Luo et al. \cite{C1_Luo} predicted new stable structures of 2D boron-carbon compounds for a wide range of boron concentrations by the PSO algorithm implemented in the CALYPSO code. XtalOpt \cite{C1_Lonie} and EVO \cite{C1_Bahmann} are the two other packages that have proved to be useful in this context. Zhu et al. proposed a method based on metadynamics and evolutionary algorithms \cite{C1_ZhuC} and found stable and metastable states for Al$_2$SiO$_5$, SiO$_2$, MgSiO$_3$, and carbon clusters \cite{C1_Qiang} from reasonable initial structures providing insight into the mechanisms of phase transitions. They designed an evolutionary algorithm based method \cite{C1_z} to automatically explore low energy surface reconstructions with variable surface atoms and reconstruction cells and illustrated it by the identification of N$_3$ trimeric reconstruction of GaN($10\bar{1}1$) surfaces. Liu \cite{C1_liu} constructed a Multi-algorithm-collaborative Universal Structure-prediction Environment (Muse) to efficiently find the stable and metastable structures of materials under given conditions. In Muse the evolutionary algorithm was coupled with the simulated annealing and the basin hopping algorithms. With the inclusion of two new variation operators, slip and twist the performance of Muse was greatly enhanced. Fadda et al. \cite{C1_Fad} described an evolutionary algorithm based on symmetry-preserving and symmetry-breaking mutations for the exploration of the space of the conjugacy classes of crystal structures. Johnston et al. have implemented and parameterized both Darwinian and Lamarkian GA search to identify structure of clusters from experimental STEM images \cite{C1_Log}.  Meredig et al. \cite{C1_yce} had given considerable theoretical and computational efforts for the development of first-principles-assisted structure solution (FPASS) that automatically solve crystal structures. Nguyen et al. \cite{C1_NJo} performed genetic algorithm to find two meta-stable Si-IX phase with good experimental agreement in the lattice parameters.

\section{Future Directions}
The evolutionary computing techniques and their hybrids have registered impressive success in the elucidation of minimum energy structures of atomic clusters and crystalline solids. These methods have not benn explored that extensively for weakly bonded molecular clusters and crystals. We anticipate rapid growth of hybrid EC based exploration of such structures. Since these surfaces are dominated by shallow minima separated by small barriers, special mutation and crossover operators would have to be designed for exhaustive exploration of the PES and locating global as well as local close-lying minima. The information on such structures and their energies could then lead to the correct prediction of thermally averaged properties of such species.

The use of EC for directly solving the molecular Schrodinger equation has been very limited. The density based method described in reference \cite{C1_sarkar} has the potential to be a viable, even a superior alternative for computing the equilibrium molecular structure and the corresponding one electron density resolved in a given basis. We anticipate extensive exploration along that line in the future years. Intelligent interfacing of the standard electronic structure codes and EC codes could prove instrumental for further growth of EC techniques in exploration of molecular structures. One area of great promise where EC could decisively prove superior is in designing molecules and materials with targetted properties. Such problems can be cast in the mould of multimodal optimization problem \cite{C1_ks2} which can be efficiently handled by the EC techniques using for example, the prey-predator model for multi-objective constrained optimization. Alongwith artificial neural network, EA can meet tate-of-the-art methods for powerful quantitative structure-property relationships modeling \cite{C1_mke,C1_le2}. We anticipate rapid growth in such studies in the next decade.
\section{Conclusions}
The last decade has seen impressive growth of EC techniques for exploring PES of clusters and crystals. The review highlights the important role played by soft-computing techniques, especially the EC methods in the elucidation of large scale structures on complex potential energy landscapes. When combined with DFT, such methods can often produce surprising results. The evolutionary computing techniques or for that matter the whole gamut of Soft-computing methods have not yet been explored exhaustively in the context of solving Schr\"odinger equation for even few electron systems although their potentials as viable tools have been demonstrated. One anticipates that the cross-talk between EC and computational chemistry would be productive and new hybrid algorithms for exploring electronic structures of atoms, molecules and clusters will evolve out of it.
\section*{Acknowledgement:}
S.P.B. thanks the DAE, Government of India, for the award of Raja Ramanna Felloship, the authorities of IIT, Bombay for hosting the fellowship, and Professor B. L. Tembe, Department of Chemistry, IIT, Bombay for many helpful discussions. We thank Dr. Pinaki Chowdhury (Calcutta University) for helpful discussions. 

\end{document}